\documentclass{article}
\usepackage{color}
\usepackage{graphicx}
\usepackage{amsmath, amsthm}
\usepackage{mathtools}
\usepackage{natbib}
\usepackage{url}
\usepackage{soulutf8}
\RequirePackage[colorlinks,citecolor=blue,urlcolor=blue]{hyperref}
\usepackage{xcolor}
\usepackage{chngcntr}
\usepackage{subfig}
\usepackage{hyperref}
\usepackage{algorithm}
\usepackage[noend]{algpseudocode}
\usepackage{setspace}

\usepackage{xfrac}
\definecolor{forest}{rgb}{0.133,0.545,0.133}
\usepackage{multirow}
\usepackage{amsfonts}

\evensidemargin=\oddsidemargin
\usepackage{etoolbox}

\newif\ifabbreviation
\pretocmd{\thebibliography}{\abbreviationfalse}{}{}
\AtBeginDocument{\abbreviationtrue}

\begin{document}
	\newcommand{\bb}{\boldsymbol{\beta}}

	\title{Bioequivalence Design with Sampling Distribution Segments}


	\author{Luke Hagar\footnote{Luke Hagar is the corresponding author and may be contacted at \url{lmhagar@uwaterloo.ca}.} \hspace{35pt} Nathaniel T. Stevens \bigskip \\ \textit{Department of Statistics \& Actuarial Science} \\ \textit{University of Waterloo, Waterloo, ON, Canada, N2L 3G1}}

	\date{}

	\maketitle

	\begin{abstract}

In bioequivalence design, power analyses dictate how much data must be collected to detect the absence of clinically important effects. Power is computed as a tail probability in the sampling distribution of the pertinent test statistics. When these test statistics cannot be constructed from pivotal quantities, their sampling distributions are approximated via repetitive, time-intensive computer simulation.  We propose a novel simulation-based method to quickly approximate the power curve for many such bioequivalence tests by efficiently exploring segments (as opposed to the entirety) of the relevant sampling distributions. Despite not estimating the entire sampling distribution, this approach prompts unbiased sample size recommendations. We illustrate this method using two-group bioequivalence tests with unequal variances and overview its broader applicability in clinical design. All methods proposed in this work can be implemented using the developed \texttt{dent} package in R.

\end{abstract}

		\bigskip


	\maketitle

	\baselineskip=19.5pt


	\section{Introduction}\label{sec:intro}

 Bioequivalence studies require a substantial investment of time, funding, and human capital. It is important to ensure these resources are invested in well-designed studies that are capable of achieving their intended objectives. Bioequivalence study objectives involve establishing the absence of a clinically important difference between treatment effects to expedite the approval process for reformulations of existing drugs. The power of a bioequivalence study is the probability of correctly establishing the absence of such effects \citep{chow2008design}. The study power generally increases with the sample size, and a power analysis is typically used to find the minimum sample size that achieves the desired power for a study.

 A power analysis considers the sampling distributions of a relevant test statistic under two hypotheses: the null hypothesis $H_0$ and alternative hypothesis $H_1$. For bioequivalence studies, $H_0$ supposes that there exists a clinically important difference between two treatments, and the alternative hypothesis $H_1$ assumes the absence of such a difference. Under the assumption that $H_0$ is true, this sampling distribution is called the null distribution. For most parametric frequentist bioequivalence tests, the null distribution coincides with a known statistical distribution based on pivotal quantities \citep{shao2003mathematical} that does not depend on the unknown model parameters. In contrast, the sampling distribution of the test statistic under $H_1$ \emph{does} depend on the magnitude of the effect size, expressed as a function of the model parameters. Power is defined as a tail probability in the sampling distribution under $H_1$, where the threshold for this tail probability is called the critical value. This tail probability is straightforward to compute via integration when the null distribution is based on a pivotal quantity, but more complex methods must be used to perform power analysis otherwise.



The null distribution is not based on pivotal quantities for many bioequivalence studies, particularly those that leverage the Welch-Satterthwaite equation \citep{satterthwaite1946approximate, welch1947generalization} to approximate the degrees of freedom for linear combinations of independent sample variances. This equation is applied to assess bioequivalence based on crossover designs \citep{lui2016crossover}, several treatments \citep{jan2020extended}, and sequential testing \citep{tartakovsky2015sequential}. Additionally, the most common application of this equation is to compare two normal population means with unequal population variances via Welch's $t$-test \citep{welch1938significance}: the default $t$-test in the programming language R. Even for this most basic use case, the null distribution is not based on a pivotal quantity. The null distribution for Welch's $t$-test approximately coincides with the standard normal distribution for large sample sizes, but this approximation based on asymptotic pivotal quantities is of limited utility since $t$-tests are most useful when the sample sizes are small (e.g., in early phases of clinical trials).

This paper presents a general framework for power analysis without the use of pivotal quantities that is primarily illustrated via two-group bioequivalence tests with unequal variances.  We focus on this setting for three reasons. First, these tests commonly assess average bioequivalence \citep{chow2008design} of two pharmaceutical drugs. Average bioequivalence compares the \emph{mean} clinical responses for two or more treatments. Second, this setting allows for clear visualization of our methodology. Third, existing methods for power analysis with these designs (see e.g., \emph{PASS} \citep{pass2023documentation}) produce unreliable results as demonstrated in this paper. While this paper emphasizes two-group bioequivalence tests with unequal variances, we later illustrate the use of the proposed methods with crossover designs. The methods are also generally applicable with equivalence, noninferiority, and one-sided hypothesis tests in nonclinical settings. The methods proposed in this paper (as well as several extensions) can all be implemented using the \texttt{dent} package in R \citep{hagar2024dent}.

Power analysis requires practitioners to choose anticipated effect sizes and variability estimates based on previous studies \citep{chow2008sample}. The recommended sample sizes achieve desired statistical power when the selected response distributions, anticipated effect sizes, and variability estimates accurately characterize the underlying data generation process. Empirical power analysis prompts more flexible methods for bioequivalence study design when the null distribution is not based on pivotal quantities. However, simulation-based methods for power analysis have computational drawbacks. Standard practice requires simulating many samples of data to reliably approximate the sampling distribution needed to estimate power for each sample size $n$ considered. This standard practice of estimating \emph{entire} sampling distributions of test statistics is wasteful because study power is a tail probability in the sampling distribution under $H_1$ defined by a critical value. It would be more computationally efficient if we could accurately assess power for a sample size $n$ by only exploring a \emph{segment} of the sampling distribution that is near the critical value. The methods for power analysis proposed in this paper adopt such an approach.


The remainder of this article is structured as follows. In Section \ref{sec:empir}, we present a method to map sampling distributions of test statistics for two-group bioequivalence tests with unequal variances to the unit cube. This mapping prompts unbiased power estimates given a pseudorandom or low-discrepancy sequence dispersed throughout the unit cube. In Section \ref{sec:curve}, we propose a novel simulation-based method that combines the mapping from Section \ref{sec:empir} with root-finding algorithms to quickly facilitate power curve approximation. This approach is fast because for a given sample size, we only explore test statistics corresponding to subspaces of the unit cube -- and hence only consider a segment of the sampling distribution. Even without estimating entire sampling distributions, we show this method yields unbiased sample size recommendations. To illustrate the wide applicability of the proposed methodology, we extend this approach for use with crossover bioequivalence designs in Section \ref{sec:cross}. Throughout the paper, we also describe how the methods can be applied with more complex study designs. We provide concluding remarks and a discussion of extensions to this work in Section \ref{sec:disc}. 




	
	\section{Mapping the Sampling Distribution to the Unit Cube}\label{sec:empir}

        \subsection{Three-Dimensional Simulation Repetitions}\label{sec:empir.reject}

        The results in this section are used to approximate power curves with segments of the relevant sampling distributions in Section \ref{sec:curve}. In this section, we describe how to map the sampling distribution of test statistics for two-group bioequivalence tests to an appropriate unit hypercube $[0,1]^w$ with low dimension $w$. This mapping allows us to implement power analyses without necessitating the high-dimensional simulation associated with repeatedly generating data. We consider a context which prompts a three-dimensional simulation corresponding to the unit cube for illustration. In particular, we suppose we collect bioavailability data $y_{ij},~ i = 1,...,n_j, ~j = 1, 2$ from the $i^{\text{th}}$ subject in group $j$. The subjects in group 1 are assigned to take the new (test) drug formulation, whereas the subjects in group 2 are assigned to take the established (reference) drug. We assume for illustration that the data $\boldsymbol{y}_j = \{y_{ij}\}_{i=1}^{n_j}$ for group $j = 1, 2$ are generated independently from a $\mathcal{N}(\mu_j, \sigma_j^2)$ distribution where $\sigma_1^2 \ne \sigma_2^2$.  Interest lies in comparing $\mu_1$ and $\mu_2$ while accounting for unequal variances.

        Bioequivalence limits $(\delta_L, \delta_U)$ are defined to determine whether differences between the formulations are clinically important. Guidance for choosing clinically important differences is often provided by regulatory bodies, such as the U.S. Food and Drug Administration (FDA) \citep{fda2003guidance, fda2022bioavailability}. However, the methods in this section accommodate any real $-\infty < \delta_L < \delta_U < \infty$. Given bioequivalence limits $\delta_L$ and $\delta_U$, we aim to conclude that $\mu_1-\mu_2 \in (\delta_L, \delta_U)$ to support average bioequivalence of the two drug formulations. We do so by rejecting the composite null hypothesis $H_0: \mu_1-\mu_2 \in (-\infty,\delta_L]\cup[\delta_U, \infty)$ in favour of the alternative hypothesis $H_1: \mu_1-\mu_2 \in (\delta_L,\delta_U)$. For such analyses, \citet{schuirmann1981hypothesis} and \citet{dannenberg1994extension} respectively proposed two one-sided test (TOST) procedures based on Student's and Welch's $t$-tests, with the Welch-based TOST procedure performing better than the standard version in the presence of unequal variances \citep{gruman2007effects, rusticus2014impact}. We henceforth refer to the Welch-based TOST procedure as the TOST procedure. For completeness, Bayesian \citep{mandallaz1981comparison, ghosh2007semi, grieve2023implementing} and nonparametric \citep{hauschke1990distribution, wellek1996new,meier2010nonparametric} alternatives to the TOST procedure for concluding average bioequivalence also exist, though this is not our focus. 
        
        The TOST procedure decomposes the interval null hypothesis $H_0$ into two one-sided hypotheses. These hypotheses are $H_{0L}: \mu_1 - \mu_2 \le \delta_L$ vs. $H_{1L}: \mu_1 - \mu_2 > \delta_L$ and $H_{0U}: \mu_1 - \mu_2 \ge \delta_U$ vs. $H_{1U}: \mu_1 - \mu_2 < \delta_U$. To conclude $\mu_1-\mu_2 \in (\delta_L, \delta_U)$, both $H_{0L}$ and $H_{0U}$ must be rejected at the nominal level of significance $\alpha$. With Welch's $t$-tests, we therefore require that
         \begin{equation*}\label{eq:test-stat}
    t_L = \dfrac{(\bar{y}_1 - \bar{y}_2) - \delta_L}{\sqrt{s_1^2/n_1  +s_2^2/n_2}} \ge t_{1 - \alpha}(\nu) ~~~ \text{and} ~~~ t_U = \dfrac{\delta_U - (\bar{y}_1 - \bar{y}_2)}{\sqrt{s_1^2/n_1  +s_2^2/n_2}} \ge t_{1 - \alpha}(\nu),
\end{equation*}
where $s_j^2$ is the sample variance for group $j= 1, 2$ and $t_{1 - \alpha}(\nu)$ is the upper $\alpha$-quantile of the $t$-distribution with $\nu$ degrees of freedom. The degrees of freedom for both $t$-tests are 
        \begin{equation}\label{eq:df}
\nu = \left(\dfrac{s_1^2}{n_1}  + \dfrac{s_2^2}{n_2}\right)^2 \times \left(\dfrac{s_1^4}{n_1^2(n_1 -1)} + \dfrac{s_2^4}{n_2^2(n_2 -1)}\right)^{-1}.
\end{equation}

Because $\nu$ is a function of the sample variances, the null distribution is not based on exact pivotal quantities. The critical value $t_{1 - \alpha}(\nu)$ for the test statistics $t_L$ and $t_U$ therefore depends on the data. The data are unknown a priori, which complicates an analytical power analysis that uses integration. \citet{jan2017optimal} considered analytical power analysis for the TOST procedure with unequal variances by expressing the test statistics in terms of simpler normal, chi-square, and beta random variables. However, the consistency of their power estimates depends on the numerical integration settings as demonstrated in Section SM2 of the online supplement. We instead use simulation to obtain consistent power estimates. To compute the test statistics $t_L$ and $t_U$, we need only simulate three sample summary statistics: $\bar{y}_1- \bar{y}_2$, $s^2_1$, and $s^2_2$. When the data are indeed generated from the anticipated $\mathcal{N}(\mu_j,\sigma_j^2)$ distributions, these sample summary statistics are sufficient and can be expressed in terms of known normal and chi-square distributions. 
    
    We generate these summary statistics using three-dimensional (3D) randomized Sobol’ sequences of length $m$: $\boldsymbol{u}_r = (u_{1r},u_{2r},u_{3r}) \in [0,1]^3$ for $r=1,...,m$. Sobol' sequences \citep{sobol1967distribution} are low-discrepancy sequences based on integer expansion in base 2 that induce negative dependence between the points $\boldsymbol{U}_1, ..., \boldsymbol{U}_m$. Sobol' sequences are regularly incorporated into quasi-Monte Carlo methods, and they can be randomized via digital shifts \citep{lemieux2009using}.  We generate and randomize Sobol' sequences in R using the \texttt{qrng} package \citep{hofert2020package}. 

When using randomized Sobol’ sequences, each point in the sequence is such that $\boldsymbol{U}_r \sim U\left([0,1]^{3}\right)$ for $r = 1, ..., m$. It follows that randomized Sobol' sequences can be used similarly to pseudorandom sequences in Monte Carlo simulation to prompt unbiased estimators:
\begin{equation}\label{eqn:qmc_exp}
    \mathbb{E}\left(\dfrac{1}{m}\sum_{r=1}^m\Psi(\boldsymbol{U}_r)\right) = \int_{[0,1]^{3}}\Psi(\boldsymbol{u})d\boldsymbol{u},
\end{equation}
for some function $\Psi(\cdot)$. Due to the negative dependence between the points, the variance of the estimator in (\ref{eqn:qmc_exp}) is typically reduced by using low-discrepancy sequences. We have that
\begin{equation}\label{eqn:qmc_var}
    \text{Var}\left(\dfrac{1}{m}\sum_{r=1}^m\Psi(\boldsymbol{U}_r)\right) = \dfrac{\text{Var}\left({\Psi(\boldsymbol{U}_r)}\right)}{m} + \dfrac{2}{m^2}\sum_{r = 1}^m\sum_{t = r + 1}^m
    \text{Cov}\left(\Psi(\boldsymbol{U}_r), \Psi(\boldsymbol{U}_t)\right),
\end{equation}
where the the first term on the right side of (\ref{eqn:qmc_var}) is the variance of the corresponding estimator based on pseudorandom sequences of independently generated points. While underutilized in simulation-based study design, low-discrepancy sequences give rise to effective variance reduction methods when the dimension of the simulation is moderate.  We can therefore use fewer simulation repetitions $m$ to obtain unbiased power estimates with Sobol' sequences in lieu of pseudorandom alternatives.

\begin{figure}[!htb] \centering 
		\includegraphics[width = \textwidth]{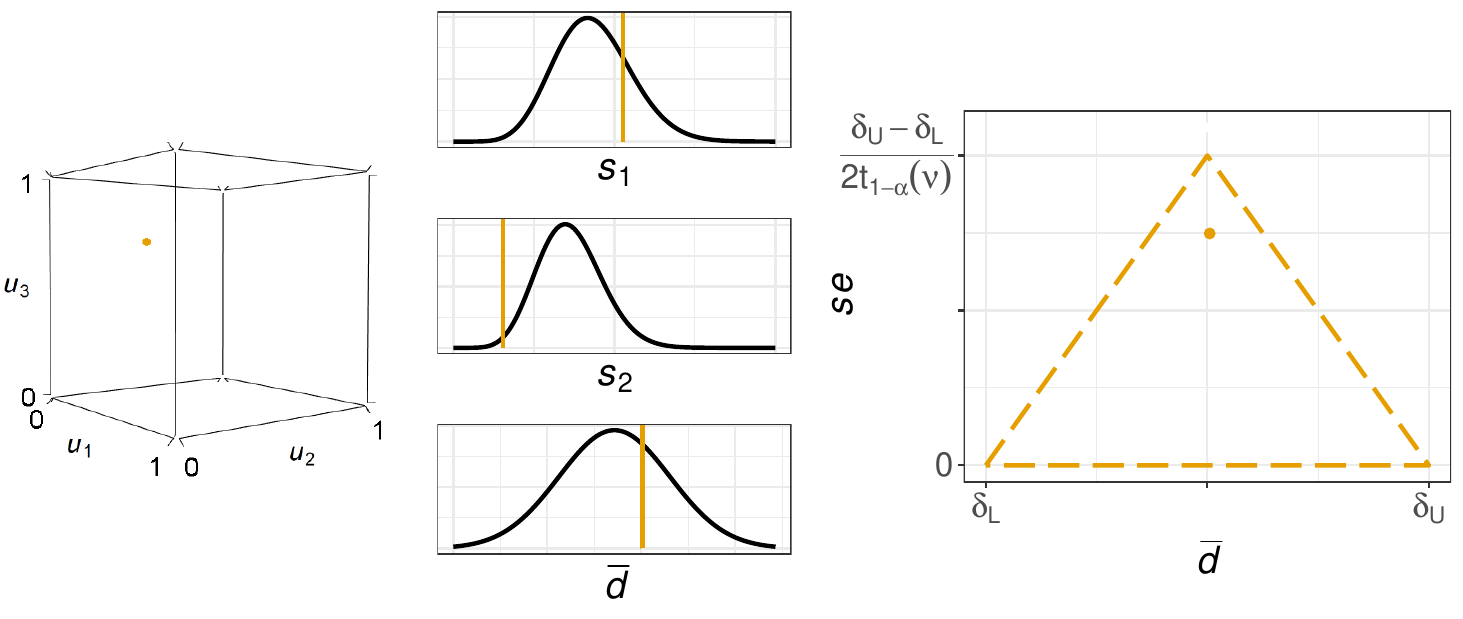} 
		\caption{\label{fig:reject_region} 
        Left: Example point $(0.785, 0.009, 0.694) \in [0,1]^3$. Center: Mapping from this point to sufficient statistics. Right: The rejection region for the TOST procedure. } 
	\end{figure}

     Algorithm \ref{alg1} outlines our procedure for unbiased empirical power estimation at sample sizes $n_1$ and $n_2$ using a Sobol’ sequence of length $m$ and significance level $\alpha$ for each $t$-test. For each of the $m$ points from the 3D Sobol' sequence, we obtain values for the summary statistics $\bar{y}_1- \bar{y}_2$, $s^2_1$, and $s^2_2$ using cumulative distribution function (CDF) inversion. We let $F(\cdot; \nu)$ and $\Phi^{-1}(\cdot)$ be the inverse CDFs of the $\chi^2_{(\nu)}$ and standard normal distributions, respectively. Given these summary statistics, we determine whether the sample for a given simulation repetition corresponds to the bioequivalence test's rejection region. The proportion of the $m$ Sobol' sequence points for which this occurs estimates the power of the test. The test statistic for each $t$-test is comprised of two random components: (1) $\bar{d} = \bar{y}_1 - \bar{y}_2$ in the numerator and (2) $se = \sqrt{s_1^2/n_1  +s_2^2/n_2}$ in the denominator. The rejection region for the TOST procedure is a triangle in the $(\bar{d}, se)$-space with vertices $(\delta_L, 0)$, $(\delta_U, 0)$, and $\left(0.5\hspace*{0.25pt}(\delta_L + \delta_U), 0.5\hspace*{0.25pt}(\delta_U - \delta_L)/t_{1 - \alpha}(\nu)\right)$. The procedure in Algorithm \ref{alg1} along with this rejection region is visualized in Figure \ref{fig:reject_region}. 
	
		\begin{algorithm}
\caption{Procedure to Compute Empirical Power}
\label{alg1}

\begin{algorithmic}[1]
\Procedure{EmpiricalPower}{$\mu_1 - \mu_2, \sigma_1, \sigma_2, \delta_L, \delta_U, \alpha, n_1, n_2, m$}
    \State \texttt{reject} $\leftarrow \textsc{null}$
    \State Generate a Sobol’ sequence of length $m$: $\boldsymbol{u}_r = (u_{1r},u_{2r},u_{3r}) \in [0,1]^3$ for $r=1,...,m$. 
    \For{$r$ in 1:$m$}
         \State Let $s_{jr}^2 = (n_{j} - 1)^{-1}\sigma_j^2 \hspace{0.25pt} F(u_{jr}; n_j - 1)$ for $j = 1, 2$
         \State Let $\bar{d}_{r} = (\mu_1 - \mu_2) + \Phi^{-1}(u_{3r}) \sqrt{\sigma_1^2/n_1 + \sigma_2^2/n_2}$ 
         \State Use $s_{1r}^2$ and $s_{2r}^2$ to compute $se_{r}$ and $\nu_r$ via (\ref{eq:df})
        \State \texttt{reject}[w] $\leftarrow$ $\texttt{ifelse(}t_{1 - \alpha}(\nu_r) \hspace{0.25pt} se_{r} < \text{min}\{ \bar{d}_{r} - \delta_L,  \delta_U - \bar{d}_{r} \}\texttt{, 1, 0)}$
    \EndFor
    \State \Return \texttt{mean(reject)} as empirical power
\EndProcedure

\end{algorithmic}
\end{algorithm}

    To apply this approach with more complex (bioequivalence) study designs, sampling distributions for hypothesis tests can be mapped to $[0,1]^{w}$, where $w$ is the number of sufficient statistics required to compute the relevant test statistics.  To design Bayesian posterior analyses, \citet{hagar2023fast} leveraged maximum likelihood estimates when low-dimensional sufficient statistics did not exist or were difficult to generate. Those methods rely on large-sample results but could be applied in frequentist settings. The simulation dimension $w$ may be large if using these mappings to design sequential tests with many interim analyses or facilitate extensive multi-group comparisons. If $w \ge 32$, caution should be exercised when using quasi-Monte Carlo methods; high-dimensional low-discrepancy sequences may have poor low-dimensional projections, which can lead to a deterioration in performance \citep{lemieux2009using}. Pseudorandom sequences could instead be used to implement such large-dimensional mappings. 

    For two-group bioequivalence tests, power analysis could be implemented by estimating power via Algorithm \ref{alg1} at various sample sizes until the desired study power of $1 - \beta$ is achieved for some type II error rate $\beta$. However, that approach would be inefficient since we would need to thoroughly explore $[0,1]^3$ -- and hence consider the entire sampling distribution -- at each combination of sample sizes considered. Low-discrepancy sequences already allow us to obtain precise power estimates using fewer points from $[0,1]^3$ than pseudorandom sequences. We can further improve this efficiency by only exploring subspaces of $[0,1]^3$ that help us estimate power. We develop a methodology for this in Section \ref{sec:curve}. But first, we introduce a motivating example that will be used to assess the performance of our method for power curve approximation proposed later. We illustrate the use of Algorithm \ref{alg1} in this context.

    \subsection{Motivating Example}\label{sec:empir.pass}

    

     This motivating example is adapted from \emph{PASS 2023} documentation \citep{pass2023documentation}. \emph{PASS} is a paid software solution that facilitates power analysis and sample size calculations for two-group (bio)equivalence tests with unequal variances. The motivating example is representative of a standard two-group bioequivalence design and seeks to compare the impact of two drugs on diastolic blood pressure, measured in mmHg (millimeters of mercury). The mean diastolic blood pressure is known to be roughly $\mu_2=96$ mmHg with the reference drug $(j=2)$, and it is hypothesized to be about $\mu_1=92$  mmHg with the test drug $(j=1)$. Subject matter experts use past studies to hypothesize within-group diastolic blood pressure standard deviations of     $\sigma_1=18$ mmHg and $\sigma_2=15$ mmHg, respectively. The bioequivalence limits for the study are $\delta_U = 0.2\times 96 = 19.2$ and $\delta_L = -\delta_U$ to comply with the FDA's $\pm20$ rule \citep{fda2006bioequivalence}. The significance level for the test is $\alpha=0.05$.
 
The \emph{PASS} documentation considers power for the motivating example at $n= n_1 = n_2 =\{3,5,8,10,15, \linebreak 20, 30,40,50,60\}$. For each sample size $n$, we estimated power 100 times using Algorithm \ref{alg1} with $m=2^{16} = 65536$. We also obtained 100 empirical power estimates for each sample size $n$ by generating $m$ samples of size $n$ from the $\mathcal{N}(\mu_1,\sigma_1^2)$ and $\mathcal{N}(\mu_2,\sigma_2^2)$ distributions and recording the proportion of samples for which we concluded that $\mu_1 - \mu_2 \in (\delta_L,\delta_U)$. Table \ref{tab:pass} summarizes these numerical results and the power estimates presented in the \emph{PASS} documentation. In recognition of their licensing agreement, \emph{PASS} software was not used nor accessed to confirm these power estimates.

\setlength{\tabcolsep}{16.4pt}

\begin{table}[tb]
\centering
\begin{tabular}{cccc} 
\hline 
   & \multicolumn{3}{c}{Estimated Power}             \\ \cline{2-4} 
$n$  & $PASS$    & Alg 1 & Na\"{\i}ve Simulation \\ \hline
3  & 0.1073 &    0.0414 ($1.43\times10^{-4}$)      &   0.0414 ($7.85\times10^{-4}$)               \\
5  & 0.1778 &      0.1283 ($1.70\times10^{-4}$)   &  0.1282 ($1.27\times10^{-3}$)                \\
8  & 0.4094  &      0.3801 ($2.60\times10^{-4}$)  &    0.3800 ($2.03\times10^{-3}$)              \\
10 & 0.5527  &     0.5366 ($2.68\times10^{-4}$)   &  0.5368 ($2.03\times10^{-3}$)                \\
15 & 0.7723  &    0.7699 ($1.49\times10^{-4}$)   &   0.7700 ($1.88\times10^{-3}$)               \\
20 & 0.8810  &    0.8815 ($1.65\times10^{-4}$)   &     0.8816 ($1.39\times10^{-3}$)             \\
30 & 0.9679  &   0.9687 ($9.32\times10^{-5}$)   &   0.9688 ($6.66\times10^{-4}$)               \\
40 & 0.9924  &    0.9922 ($5.28\times10^{-5}$)   & 0.9922 ($3.24\times10^{-4}$)                 \\
50 & 0.9982  &     0.9982 ($3.45\times10^{-5}$)    &  0.9982 ($1.67\times10^{-4}$)                \\
60 & 0.9996 &    0.9996 ($2.10\times10^{-5}$)   &    0.9996 ($7.22\times10^{-5}$)              \\ \hline
\end{tabular}
\caption{Power estimates presented in the $PASS$ documentation along with the mean of 100 empirical power estimates obtained via Algorithm \ref{alg1} (Alg 1) and simulating normal data (Na\"{\i}ve Simulation). Standard deviations of the 100 empirical power estimates are given in parentheses.}
\label{tab:pass}
\end{table}

 The two simulation-based approaches provide unbiased power estimates. However, Table \ref{tab:pass} shows that the power estimates obtained via Algorithm \ref{alg1} are much more precise than those obtained via na\"{\i}ve simulation with pseudorandom sequences. Moreover, each power estimate was obtained in roughly a quarter of a second when using Algorithm \ref{alg1}. It took between 20 and 30 seconds to obtain each power estimate using na\"{\i}ve simulation. This occurs because -- regardless of the sample size $n$ considered -- Algorithm \ref{alg1} reduces the power calculation to a three-dimensional problem that can be efficiently vectorized. We must use for loops to estimate power when directly generating the higher-dimensional data $\boldsymbol{y}_1$ and $\boldsymbol{y}_2$. Moreover, the power estimates presented in the \emph{PASS} documentation do not coincide with those returned via simulation for sample sizes less than 15, suggesting that it is valuable to develop more accurate methods for power analysis with bioequivalence designs. 

    
	
	\section{Power Curve Approximation with Segments of the Sampling Distribution}\label{sec:curve}

    \subsection{An Efficient Approach to Power Analysis}\label{sec:curve.method}

    
    In this section, we leverage the mapping between the unit cube and the test statistics presented in Section \ref{sec:empir} to facilitate power curve approximation while estimating only segments of the sampling distribution. For given sample sizes $n_1$ and $n_2$, we previously mapped each Sobol' sequence point $\boldsymbol{u}_r$ ($r = 1, ...,m$) to a mean difference, standard error, and degrees of freedom for its test statistic: $\bar{d}_{r}$, $se_{r}$, and $\nu_r$. To compute empirical power in Algorithm \ref{alg1}, we fixed the sample sizes $n_1$ and $n_2$ and allowed the Sobol' sequence point to vary. We now specify a constant $q > 0$ such that $n = n_1 = qn_2$ to allow for imbalanced sample sizes. When approximating the power curve, we fix the Sobol' sequence point $\boldsymbol{u}_r$ and let the sample size $n$ vary. We introduce the notation $\bar{d}_{r}^{_{(n,q)}}$, $se_{r}^{_{(n,q)}}$, and $\nu_r^{_{(n,q)}}$ to make this clear. For fixed $q$ and $r$, these quantities are only functions of the sample size $n$. As $n \rightarrow \infty$, $\bar{d}_{r}^{_{(n,q)}}$, $se_{r}^{_{(n,q)}}$, and $\nu_r^{_{(n,q)}}$ approach $\mu_1 - \mu_2$, 0, and $\infty$, respectively. 
    
    We consider the behaviour of these functions when $H_1$ is true, i.e., when $\mu_1 - \mu_2 \in (\delta_L, \delta_U)$. The upper vertex of the triangular rejection region for the TOST procedure is $\left(0.5\hspace*{0.25pt}(\delta_L + \delta_U), 0.5\hspace*{0.25pt}(\delta_U - \delta_L)/t_{1 - \alpha}(\nu_r^{_{(n,q)}})\right)$. First, $\nu_r^{_{(n,q)}}$ almost always increases for fixed $r$ and $q$ as $n$ increases. Thus as $n \rightarrow \infty$, the vertical coordinate of this rejection region vertex increases to $0.5\hspace*{0.25pt}(\delta_U - \delta_L)/\Phi^{-1}({1 - \alpha})$, and the remaining two vertices do not change. The rejection region then defines a threshold for the standard error $se_{r}^{_{(n,q)}}$: $\Lambda_r^{_{(n,q)}} = \text{min}\{ \bar{d}_{r}^{_{(n,q)}} - \delta_L,  \delta_U - \bar{d}_{r}^{_{(n,q)}}  \}/t_{1 - \alpha}(\nu_r^{_{(n,q)}})$. We conclude $\mu_1 - \mu_2 \in (\delta_L, \delta_U)$ if and only if $se_{r}^{_{(n,q)}}$ does not exceed this threshold. For fixed $r$ and $q$, this threshold is also a function of $n$:
\begin{equation}\label{eq:thres}
\Lambda_r^{_{(n,q)}} := \begin{cases*}
                    \dfrac{(\mu_1 - \mu_2) - \delta_L}{t_{1 - \alpha}(\nu_r^{_{(n,q)}})} + \dfrac{\Phi^{-1}(u_{3r})\sqrt{\sigma_1^2 + \sigma_2^2/q}}{\sqrt{n}\hspace*{0.25pt}t_{1 - \alpha}(\nu_r^{_{(n,q)}})} & if  $\delta_L < \bar{d}_{r}^{_{(n,q)}} \le 0.5\hspace*{0.25pt}(\delta_L + \delta_U)$  \\
                     \dfrac{\delta_U - (\mu_1 - \mu_2)}{t_{1 - \alpha}(\nu_r^{_{(n,q)}})} - \dfrac{\Phi^{-1}(u_{3r})\sqrt{\sigma_1^2 + \sigma_2^2/q}}{\sqrt{n}\hspace*{0.25pt}t_{1 - \alpha}(\nu_r^{_{(n,q)}})} & if $0.5\hspace*{0.25pt}(\delta_L + \delta_U) < \bar{d}_{r}^{_{(n,q)}} < \delta_U $ \\
                     0 & \text{otherwise.}
                 \end{cases*}
\end{equation}


    
    We suppose that a given point $\boldsymbol{u}_r$ yields $se_{r}^{_{(n,q)}} \le \Lambda_r^{_{(n,q)}}$, which corresponds to the rejection region of the TOST procedure. In Appendix \ref{sec:appendix}, we discuss why $se_{r}^{_{(n+1,q)}} \le \Lambda_r^{_{(n+1,q)}}$ generally also holds true for the same point $\boldsymbol{u}_r$. In light of this, our method to approximate the power curve generates a single Sobol' sequence of length $m$. We use root-finding algorithms \citep{brent1973algorithm} to find the smallest value of $n$ such that $se_{r}^{_{(n,q)}} \le \Lambda_r^{_{(n,q)}}$ for each point $r = 1, ..., m$. We then use the empirical CDF of these $m$ sample sizes to approximate the power curve as described in Algorithm \ref{alg2}.  


\begin{algorithm}
\caption{Procedure for Power Curve Approximation}
\label{alg2}

\begin{algorithmic}[1]
\setstretch{1}
\Procedure{PowerCurve}{$\mu_1 - \mu_2, \sigma_1, \sigma_2, \delta_L, \delta_U, \alpha$, $\beta$, $q$, $m$}
 \State \texttt{sampSobol} $\leftarrow \textsc{null}$
 \For{$r$ in 1:$m$}
 \State Generate Sobol' sequence point $\boldsymbol{u}_r$
    \State Let $\texttt{sampSobol}[r]$ solve $se_{r}^{_{(n,q)}} - \Lambda_r^{_{(n,q)}} = 0$ in terms of $n$
 \EndFor
    \State Let $n_*$ be the $(1 - \beta)$-quantile of $\texttt{sampSobol}$
    \For{$r$ in 1:$m$}
 \If{$\texttt{sampSobol}[r] \le n_*$}
 \If{$se_{r}^{_{(n,q)}} > \Lambda_r^{_{(n,q)}}$}
 \State Repeat Line 5, initializing the root-finding algorithm at $n_*$
 \EndIf
    \Else
    \If{$se_{r}^{_{(n,q)}} \le \Lambda_r^{_{(n,q)}}$}
 \State Repeat Line 5, initializing the root-finding algorithm at $n_*$
 \EndIf
    \EndIf
 \EndFor
 \State $\texttt{powerCurve} \leftarrow$ empirical CDF of $\texttt{sampSobol}$
    \State Let $n^*$ be the $(1 - \beta)$-quantile of $\texttt{sampSobol}$
 \State \Return $\texttt{powerCurve}$, $n_1 = \lceil n^* \rceil$ and $n_2 = \lceil qn^* \rceil$ as the recommended sample sizes
\EndProcedure

\end{algorithmic}
\end{algorithm}

We now elaborate on several of the steps in Algorithm \ref{alg2}. Lines 2 to 6 describe a process that would yield an unbiased power curve and sample size recommendation if $se_{r}^{_{(n,q)}} = \Lambda_r^{_{(n,q)}}$ were guaranteed to have a unique solution in terms of $n$ for fixed $r$ and $q$. However, $se_{r}^{_{(n,q)}}$ and $\Lambda_r^{_{(n,q)}}$ may infrequently intersect more than once. Given the reasoning in Appendix \ref{sec:appendix} and the numerical studies in Section SM1 of the online supplement, these multiple intersections do not occur frequently enough to deter us from using root-finding algorithms to explore sample sizes. With root-finding algorithms, we explore only subspaces of $[0,1]^3$ for each sample size investigated since different values of $n$ are considered for each point $\boldsymbol{u}_r$ in Line 4. Root-finding algorithms therefore give rise to computational efficiency as the entire sampling distribution is not estimated when exploring sample sizes $n$. In particular, the root-finding algorithm computes test statistics corresponding to $\mathcal{O}(\text{log}_2B)$ points from $[0,1]^{3}$, where $B$ is the maximum sample size considered for the power curve. We would require $\mathcal{O}(B)$ such points to explore a similar range of sample sizes using power estimates from Algorithm \ref{alg1}. When $B \ge 59$, this approach reduces the number of test statistics we must estimate by at least an order of magnitude because $\mathcal{O}(\text{log}_2B) < \mathcal{O}(B)/10$. Using low-discrepancy sequences instead of pseudorandom ones further reduces the number of test statistics we must estimate as demonstrated in Section \ref{sec:curve.ex}.

If we skipped Lines 7 to 14 of Algorithm \ref{alg2}, the unbiasedness of the sample size recommendation in Line 16 is not guaranteed due to the potential for multiple intersections between $se_{r}^{_{(n,q)}}$ and $\Lambda_r^{_{(n,q)}}$. To ensure our sample size recommendations are unbiased despite using subspaces of $[0,1]^3$ to consider sample sizes, we estimate the entire sampling distribution of test statistics at the sample size $n = n_*$ in Lines 7 to 13. If the statements in Lines 9 or 12 are true, this implies that $se_{r}^{_{(n,q)}} = \Lambda_r^{_{(n,q)}}$ for at least two distinct sample sizes $n$. For these points $\boldsymbol{u}_r$, we can reinitialize the root-finding algorithm at $n_*$ to obtain a solution for each point that will make the power curve unbiased at $n_*$. Our numerical studies in Section \ref{sec:curve.ex} show that the if statements in Lines 9 and 12 are very rarely true for any point $\boldsymbol{u}_r \in [0,1]^{3}$. In those situations, $n_* = n^*$ and both the power estimate at $n^*$ and the sample size recommendations $\lceil n^* \rceil$ and $\lceil qn^* \rceil$ are unbiased. It is incredibly unlikely that $n_*$ and $n^*$ would differ substantially, but Lines 7 to 13 of Algorithm \ref{alg2} could be repeated in that event, where the root-finding algorithm is initialized at $n^*$ instead of $n_*$. Even when $se_{r}^{_{(n,q)}}$ and $\Lambda_r^{_{(n,q)}}$ intersect more than once, the power curves from Algorithm \ref{alg2} are unbiased near the target power $1 - \beta$, but their global unbiasedness at all sample sizes $n$ is not strictly guaranteed. Nevertheless, our numerical studies in Section \ref{sec:curve.ex} highlight good global estimation of the power curve. 


The methods we leveraged to select subspaces of $[0,1]^3$ for two-group bioequivalence tests are tailored to the functions $se_{r}^{_{(n,q)}}$ and $\Lambda_r^{_{(n,q)}}$. However, these methods rely more generally on the weak law of large numbers since most sufficient statistics are based on sample means. Upon mapping the unit hypercube $[0,1]^w$ to sufficient statistics, the behaviour of the test statistics as a function of the sample size $n$ can generally be studied to develop analogs to Algorithm \ref{alg2} for other tests and designs. Root-finding algorithms are generally useful when the rejection region is convex. Rejection regions for the TOST procedure in Figure \ref{fig:reject_region}, other (bio)equivalence tests, and one-sided hypothesis tests are typically convex, whereas hypothesis tests with point null hypotheses often have non-convex rejection regions.

    \subsection{Numerical Study with the Motivating Example}\label{sec:curve.ex}

\begin{figure}[!tb] \centering 
		\includegraphics[width = 0.95\textwidth]{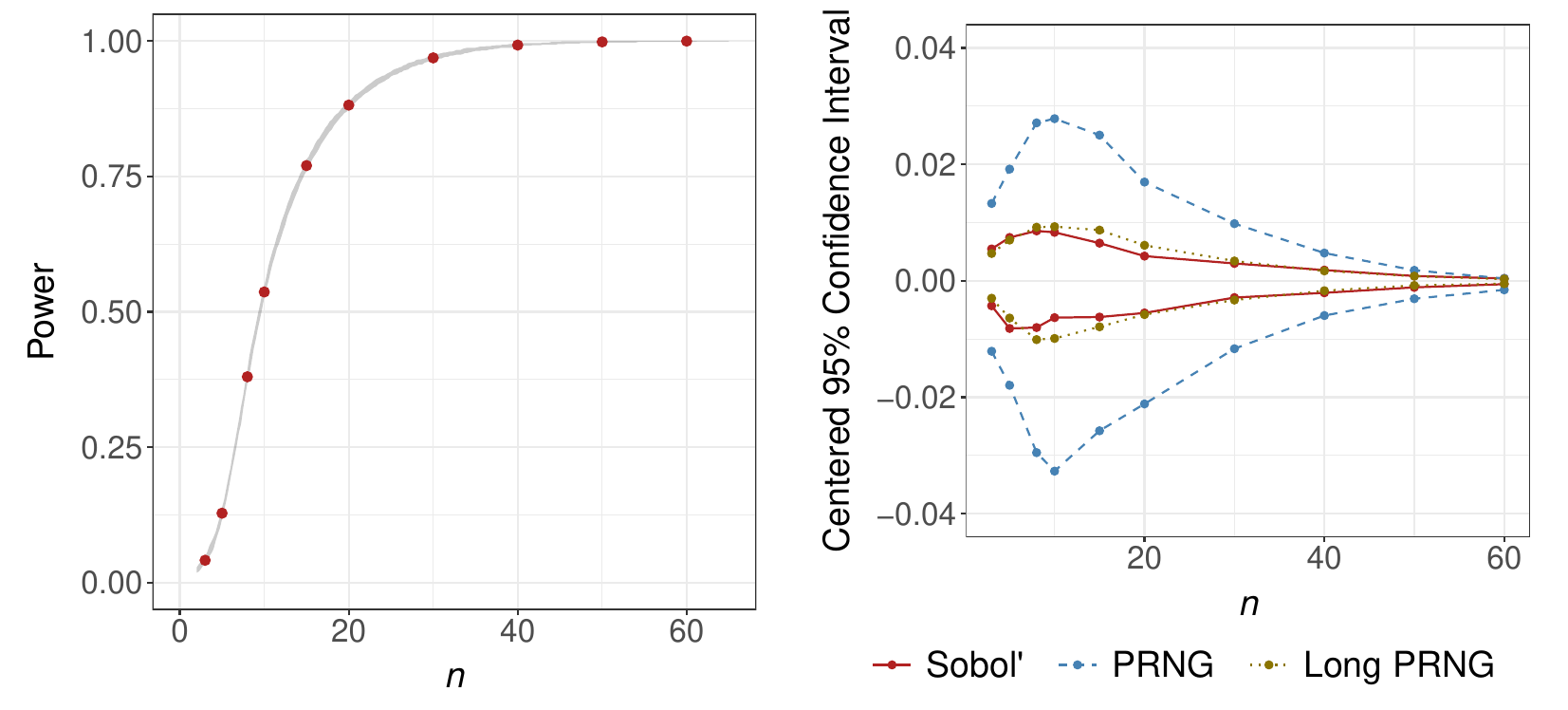} 
		\caption{\label{fig:curve} Left: 1000 power curves estimated for the motivating example (gray) and the power estimates obtained via Algorithm \ref{alg1} (red). Right: Endpoints of the centered 95\% confidence intervals for power obtained with Sobol' ($m = 1024$) and pseudorandom (PRNG) sequences ($m = 1024, 10^{4})$.} 
	\end{figure}

    We reconsider the motivating example from Section \ref{sec:empir.pass} to illustrate the reliable performance of our efficient method for power curve approximation. For this example, we approximate the power curve 1000 times with $q=1$ (i.e., $n = n_1 = n_2$). Each of the 1000 power curves are approximated using Algorithm \ref{alg2} with a target power of $1-\beta = 0.8$ and a Sobol' sequence of length $m = 2^{10} = 1024$. We recommend using shorter Sobol' sequences when approximating the power curve than when computing empirical power for a specific $(n_1, n_2)$ combination ($m = 65536$ was used in Section \ref{sec:empir.pass}). Whereas all computations in Algorithm \ref{alg1} can be vectorized, we must use a for loop to implement the root-finding algorithm for each Sobol' sequence point. We compare these 1000 power curves to the unbiased power estimates from Algorithm \ref{alg1} in Table \ref{tab:pass}. The left plot of Figure \ref{fig:curve} demonstrates that Algorithm \ref{alg2} yields suitable global power curve approximation when comparing its results to these power estimates. 

 Each power curve was approximated without estimating the entire sampling distribution for all sample sizes $n_1$ and $n_2$ explored as emphasized in Section \ref{sec:curve.target}. To further investigate the performance of Algorithm \ref{alg2}, we repeated the process from the previous paragraph to estimate 1000 power curves for the motivating example with $1 - \beta = \{0.2, 0.3,\dots, 0.7, 0.9\}$. In total, we approximated 8000 power curves for this example. Using the root-finding algorithm to explore the sample size space did not lead to performance issues. We did not need to reinitialize the root-finding algorithm in Lines 7 to 13 of Algorithm \ref{alg2} for \emph{any} of the $8.192 \times 10^6$ points used to generate these 8000 curves. The suitable performance of Algorithm \ref{alg2} is corroborated by more extensive numerical studies detailed in Section SM1 of the online supplement.

To assess the impact of using Sobol' sequences with Algorithm \ref{alg2}, we approximated 1000 power curves for the motivating example using root-finding algorithms with $1 - \beta = 0.8$ and sequences from a pseudorandom number generator. We then used the 1000 power curves corresponding to each sequence type (Sobol' and pseudorandom) to estimate power for the sample sizes considered in Section \ref{sec:empir.pass}: $n =\{3,5,8,10,15,20,30,40,$ $50,60\}$. For each sample size and sequence type, we obtained a 95\% confidence interval for power using the percentile bootstrap method \citep{efron1982jackknife}. We then created centered confidence intervals by subtracting the power estimates produced by Algorithm \ref{alg1} from each confidence interval endpoint. The right plot of Figure \ref{fig:curve} depicts these results for the 10 sample sizes $n$ and two sequence types considered. Figure \ref{fig:curve} illustrates that the Sobol' sequence gives rise to much more precise power estimates than pseudorandom sequences -- particularly when power is not near 0 or 1. We repeated this process to generate 1000 power curves via Algorithm \ref{alg2} with pseudorandom sequences of length $m = 10^4$. The power estimates obtained using Sobol' sequences with length $m = 1024$ are roughly as precise as those obtained with pseudorandom sequences of length $m = 10^4$. Using Sobol' sequences therefore allows us to estimate power with the same precision using approximately an order of magnitude fewer points from $[0,1]^3$. Each power curve for this example with $m = 1024$ took just under one second to approximate. It would take roughly 10 times as long to approximate the power curve with the same precision using pseudorandom points in lieu of Sobol' sequences.

\subsection{Exploring Subspaces of the Unit Cube}\label{sec:curve.target}

\begin{figure}[!b] \centering 
		\includegraphics[width = \textwidth]{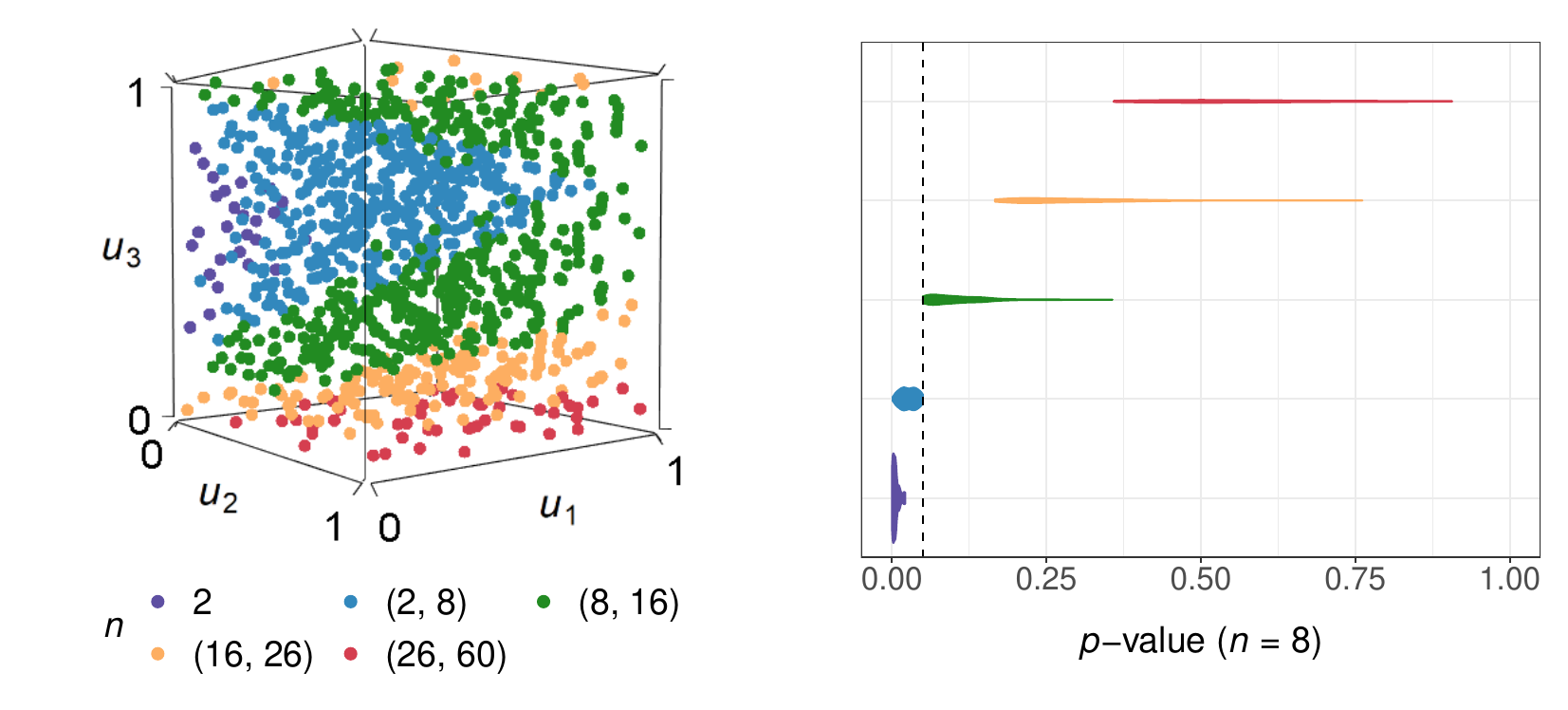} 
		\caption{\label{fig:target} Left: Visualization of which points in $[0,1]^3$ were used to explore at least one $n$ value in the various sample size ranges via the root-finding algorithm. Right: Violin plots for the sampling distribution of $p$-values corresponding to $n = 8$ conditional on the colored categorization of points. The dotted vertical line is at $\alpha = 0.05$.} 
	\end{figure}

Here, we demonstrate how segments of the sampling distribution are leveraged when exploring only subspaces of the unit cube $[0,1]^3$ for most sample sizes considered. The left plot of Figure \ref{fig:target} decomposes the results of the root-finding algorithm for one approximated power curve from Section \ref{sec:curve.ex} with $1 - \beta = 0.8$ for the motivating example. Even when the root-finding algorithm is initialized at the same sample size for all $\{\boldsymbol{u}_r \}_{r = 1}^m$, different $n$ are considered for each point $\boldsymbol{u}_r$ when determining the solution to $se_{r}^{_{(n,q)}} = \Lambda_r^{_{(n,q)}}$. The value of $n$ is noninteger in most iterations of the root-finding algorithm, and the colors in the left plot of Figure \ref{fig:target} indicate which points from the unit cube were considered for various ranges of $n$.  For instance, the purple points were such that their test statistics corresponded to the rejection region for the smallest possible sample size of $n = 2$. Moreover, only the blue points in $[0,1]^3$ were used to estimate test statistics for at least one sample size $n \in (2, 8)$ when exploring values of $n$ via the root-finding procedure.

 The points that were used to explore the smallest sample sizes generally have moderate $u_{3r}$ values and smaller $u_{1r}$ and $u_{2r}$ values. The mean difference $\bar{d}_{r}^{_{(n,q)}}$ is therefore small in absolute value and the sample variances for groups 1 and 2 are small, which implies that the numerators of the test statistics $t_L$ and $t_U$ are large and their denominators are small. The points used to explore larger sample sizes generally have more extreme $u_{3r}$ values, so $\bar{d}_{r}^{_{(n,q)}}$ may not substantially differ from one of $\delta_L$ or $\delta_U$ for small sample sizes. While the pattern in the left plot of Figure \ref{fig:target} depends on the inputs for Algorithm \ref{alg2}, the root-finding algorithm correctly identifies which subspaces of $[0,1]^3$ to prioritize for a given sample size $n$ with an arbitrary bioequivalence design. Our methods can be extended to more complex designs, but it is difficult to visualize the prioritized subspaces of the unit hypercube when the simulation dimension $w$ is greater than 3.

The right plot of Figure \ref{fig:target} visualizes the sampling distribution of $p$-values for $n = n_1 = n_2 = 8$ conditional on the categorizations from the left plot. For the TOST procedure, the $p$-value is the maximum of the $p$-values corresponding to $t_L$ and $t_U$. This $p$-value does not exceed the significance level $\alpha$ if and only if $se_{r}^{_{(n,q)}} \le \Lambda_r^{_{(n,q)}}$. This plot demonstrates why it is wasteful to use the purple points to consider $n \approx 8$ because those points satisfy $se_{r}^{_{(n,q)}} \le \Lambda_r^{_{(n,q)}}$ for $n = 2$. It follows from Appendix \ref{sec:appendix} that $se_{r}^{_{(n,q)}} \le \Lambda_r^{_{(n,q)}}$ generally holds true for $n \approx 8$ with those points, and the corresponding $p$-values are hence smaller than $\alpha = 0.05$. By similar logic, it is wasteful to consider the red points for $n \approx 8$ since the $p$-values for those points will be much larger than $\alpha$.  Although these colored categorizations are not used in Algorithm \ref{alg2}, they illustrate the targeted nature of how we consider sample sizes $n$ with segments of the relevant sampling distributions.

\section{Efficient Power Analysis for Crossover Designs}\label{sec:cross}

    The method for power curve approximation in Algorithm \ref{alg2} was tailored to a standard parallel bioequivalence study with unequal variances. However, the underlying ideas generalize to other study designs. In particular, we describe here how to extend Algorithm \ref{alg2} for use with crossover designs. Power analysis for crossover designs is of particular interest because regulatory agencies often recommend using them to conclude average bioequivalence \citep{fda2006bioequivalence}. In crossover designs, each subject receives more than one formulation of a drug during different periods \citep{chow2008design}. This is advantageous in that inter-subject variability is removed from between-formulation comparisons. We describe how to use our design methods based on sampling distribution segments with two-sequence, two-period crossover designs in Section SM3 of the online supplement. We also highlight discrepancies (of up to 33\%) between the sample sizes recommended by the power curves from Algorithm \ref{alg2} and those endorsed in popular textbooks on bioequivalence study design \citep{chow2008design}. The implementation of such extensions for these and other crossover designs is supported in the \texttt{dent} package developed in conjunction with this paper. Furthermore, our method for power analysis is flexible and could readily accommodate additional (bioequivalence) designs not discussed in this article.

	
	\section{Discussion}\label{sec:disc}

 In this paper, we developed a framework for power analysis that estimates power curves using sampling distribution segments. Our proposed methods are useful when the relevant test statistics are not based on exact pivotal quantities, which is the case for various bioequivalence designs. This framework for power analysis maps the unit hypercube $[0,1]^w$ to sufficient statistics and leverages this mapping to select sampling distribution segments. Using segments of sampling distributions improves the scalability of our simulation-based design procedures without compromising the unbiasedness of the sample size recommendations. Our framework is predominantly illustrated with three-dimensional simulation for two-group bioequivalence tests with unequal variances, but we described how to apply our methods more generally throughout the paper and now elaborate on several additional extensions.

 Future work could apply the framework proposed in this paper to design bioequivalence studies that compare more than two formulations. For such designs, the simulation dimension $w$ would need to be increased, and the multiple comparisons problem would need to be considered. It would also be important to extend this design framework to accommodate sequential analyses that allow for early termination of the (bioequivalence) study. Sequential designs consider the overall power to reject $H_0$ across all analyses: interim or final. To obtain overall power, we require the joint sampling distribution of the test statistics across all analyses to aggregate their marginal powers. In sequential settings, we would likely need to define analogues to $se_{r}^{_{(n,q)}}$ and $\Lambda_r^{_{(n,q)}}$ for each interim analysis and synthesize the results for each point $\boldsymbol{u}_r$. However, it is not trivial to create a mapping between points in $[0,1]^w$ and sufficient statistics that maintain the desired level of dependence between interim analyses for arbitrary sample sizes. Since the design of sequential tests requires that we consider sampling distributions for each planned analysis, the computational savings associated with using sampling distribution segments would be compounded in these settings.
 
 
Finally, we could explore how this framework might be applied to quickly and reliably recommend sample sizes for nonparametric bioequivalence testing methods \citep{meier2010nonparametric}. The exact null distributions for those tests are not based on pivotal quantities, and it is not possible to generate sufficient statistics in nonparametric settings. Sample size determination for these studies typically utilizes na\"{\i}ve simulation. In nonparametric settings, we may be able to map the unit hypercube $[0,1]^w$ to insufficient statistics, such as sample totals, and use low-discrepancy sequences to improve the scalability and precision of empirical power analysis. 


\section*{Supplementary Material}
These materials detail additional numerical studies, discussion of competing approaches for power analysis, and considerations for crossover designs referred to in the paper. The code to conduct the numerical studies in the paper is available online: \url{https://github.com/lmhagar/BioDesignSegments}.

	\section*{Funding Acknowledgement}
	This work was supported by the Natural Sciences and Engineering Research Council of Canada (NSERC) by way of a PGS D scholarship as well as Grant RGPIN-2019-04212.
	
	
\bibliographystyle{chicago}

\appendix

\counterwithin*{equation}{section}
\renewcommand\theequation{\thesection\arabic{equation}}

	\section{Justification for Using Root-Finding Algorithms}\label{sec:appendix}

 Here, we discuss why using root-finding algorithms to approximate the power curve yields suitable results -- even though $se_{r}^{_{(n,q)}}$ and $\Lambda_{r}^{_{(n,q)}}$ can (although infrequently) intersect more than once. The threshold $\Lambda_r^{_{(n,q)}}$ approaches $\Lambda^* = \text{min}\{ (\mu_1 - \mu_2) - \delta_L,  \delta_U - (\mu_1 - \mu_2) \}/\Phi^{-1}({1 - \alpha}) > 0$ as $n$ increases. The standard error $se_{r}^{_{(n,q)}}$ generally decreases as $n \rightarrow \infty$, but it is not necessarily a strictly decreasing function of $n$. We first consider the case where $se_{r}^{_{(n,q)}}$ does strictly decrease as $n$ increases. For small sample sizes, $\Lambda_r^{_{(n,q)}}$ is typically an increasing function of $n$ due to the decrease in $t_{1 - \alpha}(\nu_r^{_{(n,q)}})$. If the Sobol' sequence point $\boldsymbol{u}_r$ is such that $\text{sign}(u_{3r} - 0.5) = \text{sign}((\mu_1 - \mu_2) - 0.5\hspace*{0.25pt}(\delta_L + \delta_U))$, then $\Lambda_r^{_{(n,q)}}$ is also an increasing function of $n$ for large sample sizes. This occurs because $\bar{d}_{r}^{_{(n,q)}}$ is never closer than $\mu_1 - \mu_2$ to the horizontal center of the rejection region at $\bar{d} = 0.5\hspace*{0.25pt}(\delta_L + \delta_U)$. Therefore, the increasing $\Lambda_r^{_{(n,q)}}$ and decreasing $se_{r}^{_{(n,q)}}$ typically intersect once. If $\text{sign}(u_{3r} - 0.5) \ne \text{sign}((\mu_1 - \mu_2) - 0.5\hspace*{0.25pt}(\delta_L + \delta_U))$, then $\Lambda_r^{_{(n,q)}}$ is a decreasing function of $n$ for large sample sizes. This occurs because $\bar{d}_{r}^{_{(n,q)}}$ approaches $\mu_1 - \mu_2$ from the horizontal center of the rejection region. However, $\Lambda_r^{_{(n,q)}}$ decreases to a nonzero constant $\Lambda^*$, while $se_{r}^{_{(n,q)}}$ decreases to 0 as $n \rightarrow \infty$. Again, the functions $\Lambda_r^{_{(n,q)}}$ and $se_{r}^{_{(n,q)}}$ typically intersect only once.

We next consider the case where $se_{r}^{_{(n,q)}}$ is not a strictly decreasing function of $n$.  Line 5 of Algorithm \ref{alg1} prompts the first line of (\ref{eq:sdv}):
 \begin{eqnarray}\label{eq:sdv}
    se_{r}^{_{(n,q)}} &=& \dfrac{1}{\sqrt{n}} \left[ \dfrac{\sigma_1^2}{n-1} F(u_{1r}; n - 1) + \dfrac{\sigma_2^2}{q(qn-1)} F(u_{2r}; qn - 1)\right]^{1/2} \\
    &\approx& \dfrac{1}{\sqrt{2n}} \left[ \dfrac{\sigma_1^2}{n-1} \left(\Phi^{-1}(u_{1r}) + \sqrt{2(n - 1)}\right)^2 + \dfrac{\sigma_2^2}{q(qn-1)} \left(\Phi^{-1}(u_{2r}) + \sqrt{2(qn - 1)}\right)^2 \right]^{1/2}. \nonumber
\end{eqnarray}
Because quantiles from the chi-squared distribution do not have closed forms, the second line of (\ref{eq:sdv}) leverages the approximation from \citet{fisher1934statistical} for illustrative purposes. When $\Phi^{-1}(u_{1r}) \in (-\sqrt{2(n-1)} \pm 1)$ or $\Phi^{-1}(u_{2r}) \in (-\sqrt{2(qn-1)} \pm 1)$, the square function respectively makes the $(\Phi^{-1}(u_{1r}) + \sqrt{2(n - 1)})^2$ or $(\Phi^{-1}(u_{2r}) + \sqrt{2(qn - 1)})^2$ term in (\ref{eq:sdv}) smaller. As $n$ increases in those situations, the relative increase in the squared terms may offset the decreasing impact of the terms in the denominators of (\ref{eq:sdv}). However, this increasing trend cannot persist as $se_{r}^{_{(n,q)}}$ is $O\left(n^{-\frac{1}{2}}\right)$ and $Pr(\Phi^{-1}(u_{1r}) \in (-\sqrt{2(n-1)} \pm 1))$ and $Pr(\Phi^{-1}(u_{2r}) \in (-\sqrt{2(qn-1)} \pm 1))$ both approach 0 as $n \rightarrow \infty$. We show via simulation in Section SM1 of the online supplement that this increasing trend is rare for $n > 5$. For $n \le 5$, $\Lambda_{r}^{_{(n,q)}}$ is generally also an increasing function of $n$ as mentioned in Section \ref{sec:curve.method}. 

If $\Lambda_{r}^{_{(n,q)}}$ is a decreasing function of $n$, it follows from (\ref{eq:thres}) that $\bar{d}_{r}^{_{(n,q)}} = 0.5\hspace*{0.25pt}(\delta_L + \delta_U)$ when 
 \begin{eqnarray}\label{eq:n_dec}
    n = \dfrac{(\Phi^{-1}(u_{3r}))^2(\sigma_1^2 + \sigma_2^2/q)}{(0.5\hspace*{0.25pt}(\delta_L + \delta_U) - (\mu_1 - \mu_2))^2}.
\end{eqnarray}
The threshold $\Lambda_{r}^{_{(n,q)}}$ should therefore not be decreasing for sample sizes smaller than $n$ given in (\ref{eq:n_dec}). By (\ref{eq:sdv}), $se_{r}^{_{(n,q)}}$ approximates $n^{-\frac{1}{2}}\sqrt{\sigma_1^2 + \sigma_2^2/q}$ for large sample sizes $n$. It follows by (\ref{eq:thres}) that
 \begin{eqnarray}\label{eq:approx}
    \Lambda_{r}^{_{(n,q)}} \approx  \Lambda^*+ \dfrac{\lvert \Phi^{-1}(u_{3r}) \rvert}{\Phi^{-1}(1-\alpha)} se_{r}^{_{(n,q)}},
\end{eqnarray}
when $\text{sign}(u_{3r} - 0.5) \ne \text{sign}((\mu_1 - \mu_2) - 0.5\hspace*{0.25pt}(\delta_L + \delta_U))$ for large $n$.  We note that $\Lambda_{r}^{_{(n,q)}}$ and $se_{r}^{_{(n,q)}}$ may intersect for a value of $n$ that is smaller than the one given in (\ref{eq:n_dec}). If $se_{r}^{_{(n,q)}}$ is instead larger than $\Lambda_{r}^{_{(n,q)}}$ over the entire range of $n$ values for which $\Lambda_{r}^{_{(n,q)}}$ increases, then (\ref{eq:approx}) suggests that $\Lambda_{r}^{_{(n,q)}}$ and $se_{r}^{_{(n,q)}}$ are likely to intersect only once when $\Lambda_{r}^{_{(n,q)}}$ is decreasing. The functions $se_{r}^{_{(n,q)}}$ and $\Lambda_{r}^{_{(n,q)}}$ therefore typically have one intersection for all cases discussed in this appendix, but we illustrate an occurrence of multiple intersections in Section SM1 of the online supplement.

\end{document}


\newcommand{\bb}{\boldsymbol{\beta}}

	\title{Bioequivalence Design with Sampling Distribution Segments \medskip\\
 \Large{Supplementary Material}}


	\author{Luke Hagar \hspace{35pt} Nathaniel T. Stevens \bigskip \\ \textit{Department of Statistics \& Actuarial Science} \\ \textit{University of Waterloo, Waterloo, ON, Canada, N2L 3G1}}

	\date{}

	\maketitle

	\maketitle

	\baselineskip=19.5pt


 \renewcommand{\thesection}{SM\arabic{section}}

 \renewcommand{\thetable}{SM\arabic{table}}

 \renewcommand{\thefigure}{SM\arabic{figure}}


 \section{Further Justification for Using Root-Finding Algorithms}\label{sec:empir}

  \subsection{Illustration of Multiple Intersections}

 		\begin{figure}[!b] \centering 
		\includegraphics[width = 0.95\textwidth]{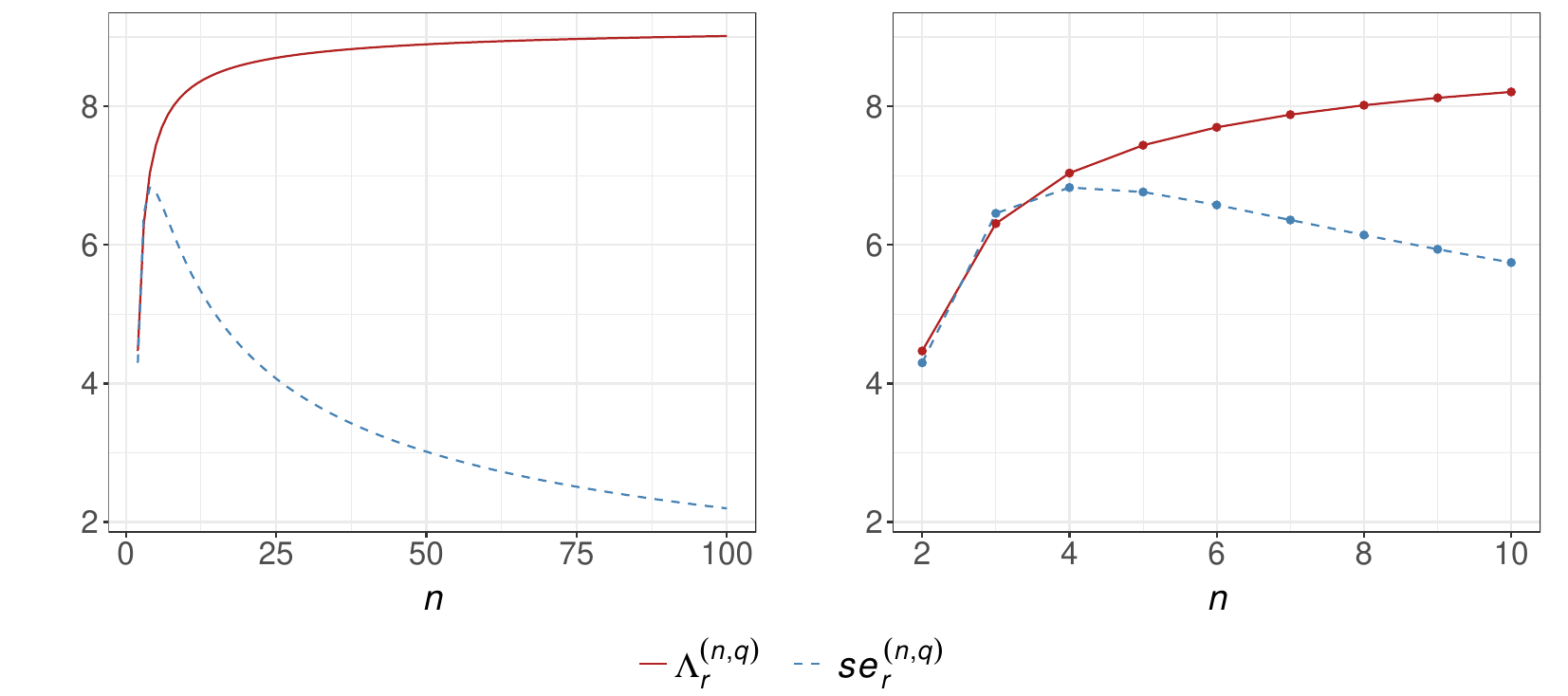} 
		\caption{\label{fig:app} Visualization of $\Lambda_{r}^{_{(n,q)}}$ and $se_{r}^{_{(n,q)}}$ as functions of $n$ for the motivating example with $\boldsymbol{u}_r = (0.184, 0.231, 0.449)$ and $q = 1$. Left: Sample sizes 2 to 100. Right: Sample sizes 2 to 10.} 
	\end{figure}
 
 We reconsider the motivating example from Section 2.2 of the main text with $q =1$. In Figure \ref{fig:app}, we show that $se_{r}^{_{(n,q)}}$ and $\Lambda_{r}^{_{(n,q)}}$ intersect more than once for the Sobol' sequence point $\boldsymbol{u}_r = (u_{1r}, u_{2r}, u_{3r}) = (0.184, 0.231, 0.449)$. We note that both $\Phi^{-1}(0.184) \in (-\sqrt{2(n-1)} \pm 1)$ and $\Phi^{-1}(0.231) \in (-\sqrt{2(n-1)} \pm 1)$ when $n = 2$; $se_{r}^{_{(n,q)}}$ is therefore small for $n =2$ and increases until $n = 4$ before decreasing to 0. This trend is evident in the right plot of Figure \ref{fig:app}, which displays the functions for sample sizes $n$ between 2 and 10. This plot shows that $se_{r}^{_{(n,q)}}$ and $\Lambda_{r}^{_{(n,q)}}$ intersect twice: once between $n = 2$ and 3 and again between $n = 3$ and 4. This means that for this point $\boldsymbol{u}_r$, $\bar{d}_{w}^{_{(n,q)}}$, $se_{r}^{_{(n,q)}}$, and $\nu_r^{_{(n,q)}}$ correspond to the rejection region of the TOST procedure for $n = 2$ and $n \ge 4$, but not for $n = 3$. The scenario visualized in Figure \ref{fig:app} arose from using a Sobol' sequence of length $m = 1024$. Of these 1024 Sobol' sequence points, there was only one other point where   $se_{r}^{_{(n,q)}}$ and $\Lambda_{r}^{_{(n,q)}}$ intersected more than once. The intersections for this other point were also between $n = 2$ and 3 and between $n = 3$ and 4.

 \subsection{The Potential for Multiple Intersections}\label{sec:empir.1}

  Here, we conduct more extensive numerical studies to justify using root-finding algorithms in our efficient approach to power curve approximation. We reconsider the motivating example from Section 2.2 of the main text, originally adapted from \emph{PASS 2023} documentation \citep{pass2023documentation}. In this example, bioavailability data were generated independently and identically for groups $j = 1$ (test) and 2 (reference) according to $\mathcal{N}(\mu_1 = 92, \sigma^2_1 = 18^2)$ and $\mathcal{N}(\mu_2 = 96, \sigma^2_2 = 15^2)$ distributions, respectively. The bioequivalence limits were $(\delta_L, \delta_U) = (-19.2, 19.2)$. The significance level for the test was $\alpha = 0.05$. We now extend this example to admit three scenarios. These three scenarios are defined by $(\sigma_1, \sigma_2) \in \{ (16.5, 16.5),  (18, 15),  (19.5, 13)\}$. We considered each scenario with $q = \{1, \sigma_2/\sigma_1, \sigma_1/\sigma_2\}$.  
  
  For each scenario and $q$ combination, we now consider sample sizes $n_1 = \{2, 3, ..., 100\}$. We require that $n_1, n_2 \ge 2$ to estimate the standard deviation for each group. For the example from Section 2.2 of the main text, $\mu_1 - \mu_2 = -4$. We also consider the motivating example where $\mu_1 - \mu_2 = \{0,-8,-12,-16\}$ with maximum $n_1$ values of $\{100, 200, 500, 2500\}$. As $\mu_1 - \mu_2$ approaches $\delta_L = -19.2$, we must consider larger sample sizes to approximate the entire power curve for those settings. Given values for $\mu_1 - \mu_2$, $q$, $\sigma_1$, and $\sigma_2$, we generated a Sobol' sequence $\boldsymbol{u}_r = (u_{1r},u_{2r},u_{3r}) \in [0,1]^3$ for $r=1,...,m$. We used $m = 1024$ for this study. For each Sobol' sequence point, we computed $se_{r}^{_{(n,q)}}$ and $\Lambda_{r}^{_{(n,q)}}$ at all $(n_1, n_2) = (n, \lfloor qn \rceil)$ pairs considered with the relevant $\mu_1 - \mu_2$ specification. We repeated this process 1000 times for each $\mu_1 - \mu_2$, $q$, $\sigma_1$, and $\sigma_2$ combination. The results from this numerical study are detailed in Table \ref{tab:mu_all}. This numerical study allows us to consider (i) scenarios where $se_{r}^{_{(n,q)}}$ and $\Lambda_{r}^{_{(n,q)}}$ intersect more than once for a given Sobol' sequence point $\boldsymbol{u}_r$ and (ii) the nondecreasing behaviour of $se_{r}^{_{(n,q)}}$ as a function of $n$. 



 \setlength{\tabcolsep}{10.1pt}

\begin{table}[!htb]
\begin{tabular}{cccccccccc}
\hline
\multicolumn{2}{c}{$\mu_1 - \mu_2 = 0$}  & & \multicolumn{3}{c}{Nonunique $se_{r}^{_{(n,q)}} = \Lambda_{r}^{_{(n,q)}}$} & & \multicolumn{3}{c}{argmax $n$ for $se_{r}^{_{(n,q)}}$}\\[0.5ex] \cline{1-2} \cline{4-6} \cline{8-10}
Scenario           & $q$  & & Prevalence & Departure & Duration & & Mean & $n> 5$ & $n > 10$ \\[0.5ex] \hline
1                  & 1 &  & 0.03\% & 3.00 & 1.04 &  & 2.54 & 2.25\% & 0.03\%       \\ \hline
\multirow{3}{*}{2} & 1 &  & 0.03\% & 3.00 & 1.04 &  & 2.55 & 2.31\% & 0.03\%       \\
                   & $1.2^{-1}$ & & 0.09\% & 3.07 & 1.12 &  & 2.89 & 3.58\% & 0.07\%       \\
                   & 1.2 & & 0.01\% & 3.00 & 1.00 &  & 2.47 & 1.83\% & 0.02\%       \\ \hline
\multirow{3}{*}{3} & 1 &  & 0.04\% & 3.00 & 1.06 &  & 2.58 & 2.58\% & 0.04\%       \\
                   & $1.5^{-1}$ & & 0.10\% & 3.01 & 1.11 &  & 2.94 & 5.84\% & 0.15\%       \\
                   & 1.5 & & 0.09\% & 3.00 & 1.01 &  & 2.48 & 2.22\% & 0.04\%       \\ \hline
\multicolumn{2}{c}{$\mu_1 - \mu_2 = -4$}  & & \multicolumn{3}{c}{Nonunique $se_{r}^{_{(n,q)}} = \Lambda_{r}^{_{(n,q)}}$} & & \multicolumn{3}{c}{argmax $n$ for $se_{r}^{_{(n,q)}}$}\\[0.5ex] \cline{1-2} \cline{4-6} \cline{8-10}
Scenario           & $q$  & & Prevalence & Departure & Duration & & Mean & $n> 5$ & $n > 10$ \\[0.5ex] \hline
1                  & 1 &  & 0.06\% & 3.00 & 1.76 &  & 2.54 & 2.25\% & 0.03\%       \\ \hline
\multirow{3}{*}{2} & 1 &  & 0.06\% & 3.01 & 1.81 &  & 2.55 & 2.31\% & 0.03\%       \\
                   & $1.2^{-1}$ & & 0.13\% & 3.31 & 1.61 &  & 2.89 & 3.58\% & 0.07\%       \\
                   & 1.2 & & 0.05\% & 3.00 & 1.71 &  & 2.47 & 1.82\% & 0.02\%       \\ \hline
                   
\multirow{3}{*}{3} & 1 &  & 0.07\% & 3.01 & 1.78 &  & 2.58 & 2.58\% & 0.04\%       \\
                   & $1.5^{-1}$ & & 0.14\% & 3.31 & 1.65 &  & 2.94 & 5.84\% & 0.15\%       \\
                   & 1.5 & & 0.16\% & 3.00 & 1.34 &  & 2.48 & 2.21\% & 0.04\%       \\ \hline
\multicolumn{2}{c}{$\mu_1 - \mu_2 = -8$}  & & \multicolumn{3}{c}{Nonunique $se_{r}^{_{(n,q)}} = \Lambda_{r}^{_{(n,q)}}$} & & \multicolumn{3}{c}{argmax $n$ for $se_{r}^{_{(n,q)}}$}\\[0.5ex] \cline{1-2} \cline{4-6} \cline{8-10}
Scenario           & $q$  & & Prevalence & Departure & Duration & & Mean & $n> 5$ & $n > 10$ \\[0.5ex] \hline
1                  & 1 &  & 0.16\% & 3.14 & 3.35 &  & 2.54 & 2.25\% & 0.03\%       \\ \hline
\multirow{3}{*}{2} & 1 &  & 0.17\% & 3.13 & 3.42 &  & 2.55 & 2.31\% & 0.04\%       \\
                   & $1.2^{-1}$ & & 0.23\% & 3.76 & 3.18 &  & 2.89 & 3.58\% & 0.07\%       \\
                   & 1.2 & & 0.15\% & 3.12 & 3.27 &  & 2.47 & 1.82\% & 0.02\%       \\ \hline
\multirow{3}{*}{3} & 1 &  & 0.18\% & 3.14 & 3.42 &  & 2.58 & 2.56\% & 0.04\%       \\
                   & $1.5^{-1}$ & & 0.31\% & 4.08 & 3.07 &  & 2.94 & 5.84\% & 0.16\%       \\
                   & 1.5 & & 0.33\% & 3.05 & 2.42 &  & 2.48 & 2.22\% & 0.03\%       \\ \hline
\multicolumn{2}{c}{$\mu_1 - \mu_2 = -12$}  & & \multicolumn{3}{c}{Nonunique $se_{r}^{_{(n,q)}} = \Lambda_{r}^{_{(n,q)}}$} & & \multicolumn{3}{c}{argmax $n$ for $se_{r}^{_{(n,q)}}$}\\[0.5ex] \cline{1-2} \cline{4-6} \cline{8-10}
Scenario           & $q$  & & Prevalence & Departure & Duration & & Mean & $n> 5$ & $n > 10$ \\[0.5ex] \hline
1                  & 1 &  & 0.36\% & 3.46 & 8.03 &  & 2.54 & 2.25\% & 0.03\%       \\ \hline
\multirow{3}{*}{2} & 1 &  & 0.38\% & 3.46 & 8.15 &  & 2.55 & 2.32\% & 0.03\%       \\
                   & $1.2^{-1}$ & & 0.49\% & 4.47 & 7.78 &  & 2.89 & 3.58\% & 0.07\%       \\
                   & 1.2 & & 0.38\% & 3.39 & 7.34 &  & 2.47 & 1.82\% & 0.02\%       \\ \hline
\multirow{3}{*}{3} & 1 &  & 0.41\% & 3.48 & 8.02 &  & 2.58 & 2.57\% & 0.04\%       \\
                   & $1.5^{-1}$ & & 0.65\% & 5.06 & 6.88 &  & 2.94 & 5.84\% & 0.15\%       \\
                   & 1.5 & & 0.60\% & 3.26 & 5.89 &  & 2.48 & 2.22\% & 0.04\%       \\ \hline
\multicolumn{2}{c}{$\mu_1 - \mu_2 = -16$}  & & \multicolumn{3}{c}{Nonunique $se_{r}^{_{(n,q)}} = \Lambda_{r}^{_{(n,q)}}$} & & \multicolumn{3}{c}{argmax $n$ for $se_{r}^{_{(n,q)}}$}\\[0.5ex] \cline{1-2} \cline{4-6} \cline{8-10}
Scenario           & $q$  & & Prevalence & Departure & Duration & & Mean & $n> 5$ & $n > 10$ \\[0.5ex] \hline
1                  & 1 &  & 0.77\% & 4.33 & 35.72 &  & 2.54 & 2.25\% & 0.04\%       \\ \hline
\multirow{3}{*}{2} & 1 &  & 0.79\% & 4.37 & 35.38 &  & 2.55 & 2.31\% & 0.04\%       \\
                   & $1.2^{-1}$ & & 1.01\% & 5.79 & 33.17 &  & 2.89 & 3.58\% & 0.07\%       \\
                   & 1.2 & & 0.84\% & 4.24 & 31.50 &  & 2.47 & 1.82\% & 0.02\%       \\ \hline
\multirow{3}{*}{3} & 1 &  & 0.86\% & 4.36 & 35.54 &  & 2.58 & 2.57\% & 0.05\%       \\
                   & $1.5^{-1}$ & & 1.24\% & 6.66 & 29.49 &  & 2.94 & 5.85\% & 0.15\%       \\
                   & 1.5 & & 1.20\% & 4.22 & 25.87 &  & 2.48 & 2.22\% & 0.04\%       \\ \hline
\end{tabular}
\caption{Simulation results for 1000 repetitions of all scenario and $q$ combinations when $\mu_1 - \mu_2 = \{0, -4, -8, -12, -16\}$ with $m = 1024$. The center section of the table concerns scenarios where $se_{r}^{_{(n,q)}} = \Lambda_{r}^{_{(n,q)}}$ have a nonunique solution. The right section of the table concerns nondecreasing behaviour of $se_{r}^{_{(n,q)}}$.}
\label{tab:mu_all}
\end{table}

The center section of Table \ref{tab:mu_all} concerns scenarios where $se_{r}^{_{(n,q)}} = \Lambda_{r}^{_{(n,q)}}$ have nonunique solutions. The prevalence column indicates the mean percentage of the $m = 1024$ Sobol' sequence points that had multiple solutions for $se_{r}^{_{(n,q)}} = \Lambda_{r}^{_{(n,q)}}$. This percentage is very low, particularly when $\mu_1 - \mu_2$ is close to the center of the equivalence region $0.5\hspace{0.25pt}(\delta_U - \delta_L) = 0$. The prevalence of multiple intersections increases as $\mu_1 - \mu_2$ approaches $\delta_L = -19.2$, but $se_{r}^{_{(n,q)}}$ and $ \Lambda_{r}^{_{(n,q)}}$ intersect only once for roughly 99\% of the Sobol' sequence points when $\mu_1 - \mu_2 = -16$. For the Sobol' sequence points with nonunique solutions, the departure column details the mean value of $n$ such that $\widetilde{se}_{r}^{_{(n-1,q)}} < \Lambda_{r}^{_{(n-1,q)}}$ but $se_{r}^{_{(n,q)}} > \Lambda_{r}^{_{(n,q)}}$. That is, this column summarizes the mean value at which this Sobol' sequence point leaves the rejection region for the TOST procedure. This sample size is very small for all scenarios considered. In the vast majority of situations, this departure occurs at a sample size of 3 (i.e., $\boldsymbol{u}_r$ prompts a sample that is in the rejection region when $n = 2$ but not when $n = 3$). 



The duration column summarizes the mean value for the smallest $\zeta \in \mathbb{Z}^+$ such that $\widetilde{se}_{r}^{_{(n + \zeta,q)}} < \Lambda_{r}^{_{(n + \zeta,q)}}$ for the departing sample size $n$ (i.e., the number of sample sizes before the sample corresponding to $\boldsymbol{u}_r$ returns to the TOST rejection region). The mean duration of these departures increases as $\mu_1 - \mu_2$ approaches $\delta_L = - 19.2$ but so do the sample sizes $n_1$ and $n_2$ required to achieve the desired target power. For instance, we require $n$ between roughly 200 and 450 to obtain 80\% power for the settings where $\mu_1 - \mu_2 = -16$. Therefore, the mean duration of these departures is small with respect to the recommended sample sizes. 

The right section of Table \ref{tab:mu_all} concerns the nondecreasing behaviour of $se_{r}^{_{(n,q)}}$, which does not depend on the value for $\mu_1 - \mu_2$. The mean column indicates the average sample size $n$ at which $se_{r}^{_{(n,q)}}$ peaks over all simulation repetitions. Because a minimum $n$ value of 2 is required to estimate $\sigma_1$ and $\sigma_2$, $se_{r}^{_{(n,q)}}$ is a generally decreasing function of $n$ for the majority of Sobol' sequence points. As indicated in the two rightmost columns of Table \ref{tab:mu_all}, it is uncommon for $se_{r}^{_{(n,q)}}$ to peak at sample sizes $n > 5$, and nondecreasing behaviour of $se_{r}^{_{(n,q)}}$ is incredibly rare for $n > 10$. These results are encouraging because the nondecreasing behaviour of $se_{r}^{_{(n,q)}}$ drives many of the multiple intersections between $se_{r}^{_{(n,q)}}$ and $\Lambda_{r}^{_{(n,q)}}$. 



 

 \subsection{The Impact of Multiple Intersections on Power Curve Approximation}

 As visualized in Section 3.3 of the main text, Algorithm 2 approximates power curves without estimating the entire sampling distribution of test statistics for all sample sizes. The segments of the relevant sampling distributions are selected using root-finding algorithms under the assumption that the functions $se_{r}^{_{(n,q)}}$ and $\Lambda_{r}^{_{(n,q)}}$ have a unique solution for each point $\boldsymbol{u}_r, r = 1, ..., m$. Reinitializing the root-finding algorithm in Lines 7 to 13 of Algorithm 2 allows us to obtain an unbiased sample size recommendation in the presence of multiple intersections. In Section 3.2 of the main text, we conducted a numerical study with 8000 power curves for the motivating example in which the root-finding algorithm never needed to be reinitialized. We now extend that numerical study to the expanded set of scenarios defined in Section \ref{sec:empir.1}. 

 We considered the 35 scenarios from Table \ref{tab:mu_all}. These scenarios detail five values for $\mu_1 - \mu_2$: $\{0,-4,-8,\linebreak-12,-16\}$. For each $\mu_1 - \mu_2$ value, seven $(\sigma_1, \sigma_2, q)$ combinations explored bioequivalence with unequal variances and imbalanced sample sizes: $\{1 = (16.5, 16.5, 1),~ 2 = (18, 15, 1),~ 3 = (18, 15, 1.2^{-1}),~ 4 = (18, 15, 1.2),\linebreak~ 5 = (19.5, 13, 1),~ 6 = (19.5, 13, 1.5^{-1}),~ 7 = (19.5, 13, 1.5) \}$. For each of these 35 scenarios, we considered eight values for the target power $1 - \beta = \{0.2, 0.3,\dots, 0.9\}$. The remaining inputs for Algorithm 2 are $\alpha = 0.05$, $\delta_L = - 19.2$, $\delta_U = 19.2$, and $m = 1024$ as used in Section 3.2. For each of the $35\times8 = 280$ scenario and target power combinations, we approximated 100 power curves using Algorithm 2. 
 
 We only needed to reinitialize the root-finding algorithm in Lines 7 to 13 of Algorithm 2 for four of the $2.867 \times 10^7$ points used to generate these 28000 curves. Those four points were used for scenarios where $1- \beta = 0.2$ and $\mu_1 - \mu = - 12$; two of those points were used with the first $(\sigma_1, \sigma_2, q)$ combination, and one of those points was used with each of the third and fifth $(\sigma_1, \sigma_2, q)$ combinations. Those four points prompted multiple intersections between $se_{r}^{_{(n,q)}}$ and $\Lambda_{r}^{_{(n,q)}}$, one of which occurred before $n_*$ in Line 6 of Algorithm 2 and the other of which occurred after. We needed to choose the other intersection to obtain an unbiased power estimate -- even though $\lceil n_* \rceil$ and $\lceil n^* \rceil$ from Algorithm 2 were the same for the four power curves created using these points. We therefore very rarely need to adjust for multiple intersections, especially for high-powered studies since the root-finding algorithm never needed to be reimplemented for $1 - \beta \ge 0.3$. However, we cannot guarantee that it is unnecessary to adjust for multiple intersections with an arbitrary (bioequivalence) design, so that is why we incorporated Lines 7 to 13 into Algorithm 2 to ensure unbiased sample size recommendations.


 

  \section{Competing Methods for Power Analysis}



In this section, we consider alternative methods for power analysis with the Welch-based TOST procedure. We illustrate that the consistency of the power estimates produced by such methods depends on the numerical integration settings. For the Welch-based TOST procedure, \citet{jan2017optimal} proposed an analytical method to compute power given sample sizes $n_1$ and $n_2$. They accounted for the degrees of freedom being unknown a priori by expressing the test statistic in terms of simpler normal, chi-square, and beta random variables. The sum of the sample variances for the two groups is related to a chi-square distribution, and the proportion of total variability arising from the first group is related to a beta distribution. Power was computed by integrating with respect to the expectation of these (independent) chi-square and beta random variables. \citet{shieh2022exact} provided R code to implement exact power calculations for the Welch-based TOST procedure using two-dimensional quadrature with Simpson's rule \citep{suli2003introduction}. 

We computed power estimates using \citeauthor{jan2017optimal}'s (\citeyear{jan2017optimal}) method for the motivating example from Section 2.2 of the main text with $n = \{3,5,8,10,15, 20, 30,40,50,60\}$.
When the numerical integration parameters for Simpson's rule are properly tuned, these power estimates coincided with those obtained by Algorithm 1 in Table 1 of the main text to four decimal places. With $n = 2$, \citeauthor{jan2017optimal}'s (\citeyear{jan2017optimal}) method provided a power estimate of 2.0838 using the recommended quadrature settings with $5\times10^5$ points. When using the default settings with $5\times10^6$, $5\times10^7$, and $2.5\times10^8$ points, their method respectively estimated power to be 0.2296, 0.0443, and 0.0278. The final estimate took roughly 56 seconds to compute on a standard laptop. For $n = 2$, we estimated power 100 times using Algorithm 1 with $m = 65536$ as done in Table 1. This gave rise to an empirical power estimate of 0.0238 and a 95\% confidence interval of $(0.0236, 0.0240)$ created using the percentile bootstrap method \citep{efron1982jackknife}; this confidence interval does not contain the final estimate returned by \citeauthor{jan2017optimal}'s (\citeyear{jan2017optimal}) method of 0.0278. 

We also note that when fewer than $5\times10^7$ points are used with their default settings, power is not a strictly nondecreasing function of the sample size $n$ for the motivating example. This occurs because the quadrature rule has not converged. This lack of convergence is problematic -- even for the smallest possible sample size of $n = 2$. To find a suitable sample size, power is often computed successively for $n = \{2,3,4, ...\}$ until the target power $1 - \beta$ for the study is achieved. If sample sizes are explored using a bisection method, this process is initialized by computing power at lower and upper bounds for $n$. This lower bound is typically $n = 2$. Depending on the chosen quadrature settings, using either approach for this example could lead us to incorrectly conclude that a sample size of $n = 2$ is sufficient for a high-powered bioequivalence study.

\begin{figure}[!htb] \centering 
		\includegraphics[width = 0.95\textwidth]{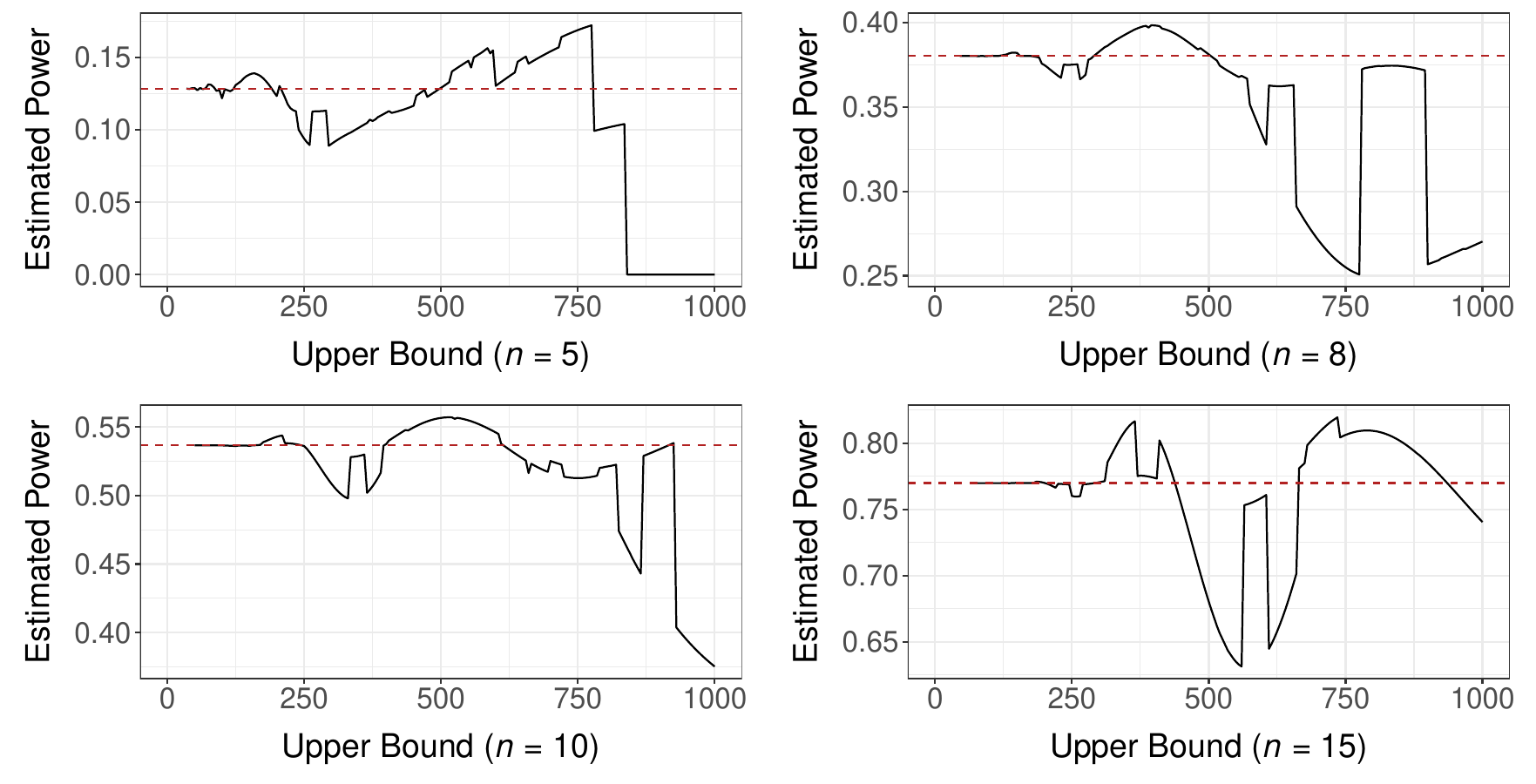} 
		\caption{\label{fig:pracma} Estimated power (black) for the motivating example with $n = \{5, 8, 10, 15\}$ and various upper bounds of integration for the chi-square variable. Actual power for these designs is visualized in red.} 
	\end{figure}

Alternative numerical integration techniques may also yield unstable results when combined with \citeauthor{jan2017optimal}'s (\citeyear{jan2017optimal}) method. We modify \citeauthor{shieh2022exact}'s (\citeyear{shieh2022exact}) code to compute power using two-dimensional numerical integration techniques in R. This requires us to integrate over a beta variable with domain $(0,1)$ and a chi-square variable with domain $\mathbb{R}^+$. In practice, we often need to choose a finite upper bound of integration for the chi-square variable. Figure \ref{fig:pracma} illustrates that the estimated power for the motivating example at various sample sizes $n$ is sensitive to this choice of upper bound when implementing numerical integration via R's \texttt{pracma} package \citep{borchers2021pracma}.


Thus, the consistency of the power estimates returned by competing methods depends on the chosen integration bounds or point grid for the quadrature rule. These issues with consistency can be diagnosed when considering various values for the numerical integration parameters, but these considerations might not be made when exploring power at different sample sizes using an automated process. \citeauthor{jan2017optimal}'s (\citeyear{jan2017optimal}) method computes power for fixed sample sizes $n_1$ and $n_2$, so it is comparable to Algorithm 1 from the main text. Algorithm 1's equivalent of the numerical integration parameters is the length $m$ of the Sobol' sequence. We emphasize that this choice for $m$ only impacts the precision -- and not the consistency -- of the power estimates.

	\section{Power Analysis for Crossover Designs}\label{sec:cross}


While certain studies assess average bioequivalence via parallel designs, the FDA and other regulatory agencies often recommend that crossover designs be used instead. In crossover designs, each subject receives more than one formulation of a drug during different periods \citep{chow2008design}. Each group (or block) of subjects receives a different sequence of formulations. Crossover designs possess several advantages over their parallel counterparts. First, each subject can serve as their own control, which facilitates within-subject comparison between drug formulations. Crossover designs also remove inter-subject variability from between-formulation comparisons.

Although crossover designs often require fewer subjects to obtain desired power for the equivalence test, they take longer to implement than parallel designs because each subject is analyzed over multiple treatment periods. Moreover, there are typically rest periods between consecutive treatment periods so that the effect of the formulation administered in one treatment period does not persist in the next. These rest periods are called \emph{washout} periods, and they should be long enough for the effect of one formulation to wear off so that there is no \emph{carryover} effect in the next treatment period. If the washout period length is too short relative to the persistence of the formulation effects, we must distinguish between the effect of the drug being administered in a given period (direct drug effect) and the carryover effect. First-order carryover effects are those that last a single treatment period. Generally, higher-order carryover effects that last multiple treatment periods are not considered in bioequivalence studies.

There are many crossover designs that assess average bioequivalence, the most common of which is the two-sequence, two-period ($2 \times 2$) crossover design. In the $2 \times 2$ crossover design, subjects are assigned to sequence 1 (RT) or 2 (TR). The acronyms in parentheses denote which order the subjects in that sequence receive the test and reference formulations. There is a washout period between the two treatment periods. We consider the statistical model for the $2 \times 2$ crossover design described in \citet{chow2008design}. We let $Y_{ijk}$ be the response from the $i^{\text{th}}$ subject in the $k^{\text{th}}$ sequence at the $j^{\text{th}}$ period such that
         \begin{equation}\label{eq:cross.model1}
    Y_{ijk} = \mu + S_{ik} + P_j + F_{(j,k)} + C_{(j-1,k)} + e_{ijk},
\end{equation}
where $i = 1, ..., n_k$, $j = 1,2$, and $k = 1,2$. Here, $n_k$ is the number of subjects in the $k^{\text{th}}$ sequence, and $\mu$ is the overall mean. $S_{ik}$ is the random effect for the $i^{\text{th}}$ subject in the $k^{\text{th}}$ sequence, and we assume that these terms are independently and identically distributed (i.i.d.) according to a normal distribution with mean 0 and variance $\sigma^2_S$. $P_j$ is the fixed effect of the $j^{\text{th}}$ period such that $P_1 + P_2 = 0$. $F_{(j,k)}$ is the direct fixed effect of the formulation administered to subjects in sequence $k$ during the $j^{\text{th}}$ period. For the $2 \times 2$ crossover design, we have that $F_{(j,k)} = F_R$ if $j = k$ and $F_T$ otherwise. We assume that $F_T + F_R = 0$. $C_{(j-1,k)}$ is the fixed first-order carryover effect of the formulation administered in the $(j-1)^{\text{th}}$ period of sequence $k$, where $C_{(0,k)} = 0$ for $k = 1,2$. Furthermore, we have that $C_{(1,1)} = C_R$, $C_{(1,2)} = C_T$, and $C_T + C_R = 0$. Finally, $e_{ijk}$ is the within-subject random error, where these terms are assumed to be i.i.d.\ normal with mean 0 and variance $\sigma^2_T$ or $\sigma^2_R$ depending on the formulation administered. We further assume that the $S_{ik}$ and $e_{ijk}$ terms are mutually independent.

The $2 \times 2$ crossover design allows us to consider the presence of carryover effects by testing the hypothesis that $C_T - C_R = 0$ \citep{chow2008design}. However, we cannot uniquely estimate model (\ref{eq:cross.model1}) using a $2 \times2$ crossover design if carryover effects are present. In such scenarios, we also cannot obtain an unbiased estimator for the direct drug effect $F = F_T - F_R$ based on data from both periods. If carryover effects are present, only the data from the first period is typically used. This effectively reverts the design into a parallel study, and the methods from the main text can be applied. Here, we assume the absence of carryover effects (i.e., $C_T = C_R = 0$) and consider power analysis with carryover effects in the \texttt{dent} package. We define period differences for each subject within each sequence as
         \begin{equation}\label{eq:intra.diff}
    D_{ik} = \frac{1}{2}(Y_{i2k} - Y_{i1k}),
\end{equation}
for $i = 1, 2, ..., n_k$ and $k = 1,2$. In the absence of carryover effects, an unbiased estimator for the direct drug effect is $\hat F = \bar{D}_{\cdot 1} - \bar{D}_{\cdot 2}$. Under model (\ref{eq:cross.model1}), $\sigma^2_{Dk} = Var(D_{ik}) = (\sigma^2_T + \sigma^2_R)/4$ for both sequences $k = 1,2$, and the equal variance assumption is theoretically sound. However, this assumption may be inappropriate if we allow the within-subject random errors to vary by treatment and sequence (i.e., $Var(e_{ijk})$ is $\sigma^2_{Tk}$ or $\sigma^2_{Rk}$ depending on the formulation administered in period $j$ of sequence $k$). The equal variance assumption may also be inappropriate for crossover designs that account for carryover effects, such as two-sequence dual designs \citep{chow2008design}. 

It is therefore useful to have Welch-based design methods for crossover studies that allow for unequal variances. Algorithms 1 and 2 from the main text can readily be extended to serve this purpose. For parallel designs, an anticipated value for the drug effect $\mu_1 - \mu_2$ is chosen. In the $2 \times 2$ crossover design, a similar input for the sample size calculation is specified for $F = F_T - F_R$. Instead of hypothesizing values for inter-subject standard deviations $\sigma_1$ and $\sigma_2$, practitioners guess values for the standard deviations of the intra-subject differences in sequences 1 and 2: $\sigma_{D1}$ and $\sigma_{D2}$. Algorithms 1 and 2 can be directly applied with the $2 \times 2$ crossover design by substituting $\mu_1 - \mu_2$ with $F$, $\sigma_1$ with $\sigma_{D1}/2$, and $\sigma_2$ with $\sigma_{D2}/2$. The intra-subject standard deviations are divided by two due to the factor of $1/2$ in (\ref{eq:intra.diff}). 


We illustrate the value of this approach using an example from \citet{chow2008sample}. This example concerns a clinical trial that assesses average bioequivalence between a test formulation of a drug that is inhaled and a reference formulation that is injected. The formulations are compared using log-transformed area under the curve (AUC), which measures total drug exposure across time in pharmacokinetics contexts. The bioavailability data are assumed to be normal after this logarithmic transformation. The mean difference of AUC is assumed to be $F = 0.05$. The bioequivalence limits are chosen to be $\delta_U = -\delta_L = 0.223$ to comply with FDA requirements \citep{fda2003guidance}. Balanced samples are to be collected ($n = n_1 = n_2$), and past studies give rise to anticipated intra-subject standard deviations of $\sigma_{D1} = \sigma_{D2} = \sigma_D = 0.4$. The investigator wants to find the sample size $n$ that achieves $100\times(1-\beta)\% = 80\%$ power at the significance level of $\alpha = 0.05$. \citet{chow2008sample} recommended conservatively choosing the smallest sample size $n$ that satisfies
         \begin{equation}\label{eq:chow.n}
    n \ge \dfrac{(t_{\alpha, 2n-2} + t_{\beta/2, 2n-2})^2\sigma_D^2}{2(\delta_U - \lvert F \rvert)^2}.
\end{equation}

Because (\ref{eq:chow.n}) does not admit an explicit solution for the sample size per sequence, the desired $n$ must be found numerically. As such, tables populated with $n$ values corresponding to various $F$, $\beta$, and $\sigma_D$ combinations are often used to select sample sizes. One such table in \citet{chow2008sample} recommended a sample size of $n = 24$ per sequence with this example. We first implemented Algorithm 2 from the main text for unequal variances with $\sigma_{D1} = \sigma_{D2} = 0.4$. For comparison, we also implemented an equal variance version of this approach that uses two-dimensional Sobol' sequences and the TOST procedure with Student's $t$-tests. Both the equal and unequal variance versions of our approach return a recommended sample size of $n = 18$. Given that $ n_1 = n_2$ and $\sigma_{D1} = \sigma_{D2}$, it is not surprising that both approaches recommend the same sample size based on numerical studies from \citet{gruman2007effects} and \citet{rusticus2014impact}. 

A crossover study with $2 \times 24 = 48$ total subjects takes substantially more resources to conduct than one with only 36 subjects. \citet{chow2008sample} acknowledged that (\ref{eq:chow.n}) returns a conservative sample size, but the degree of conservatism is not transparent. Their sample size recommendation is conservative when $\delta_U - \lvert F \rvert < \lvert \delta_L \rvert$ as in this example. For this example, using (\ref{eq:chow.n}) to choose a sample size effectively changes the lower bioequivalence limit to $\delta_L = F - (\delta_U - F) = -0.123$. Both versions of Algorithm 2 with equal and unequal variances align with (\ref{eq:chow.n}) and recommend $n = 24$ when $(\delta_L, \delta_U) = (-0.123, 0.223)$. Our approaches therefore better accommodate scenarios where $F \ne 0.5\hspace{0.25pt}(\delta_L + \delta_U)$ than certain design methods that leverage static tables or analytical formulas -- even when the equal variance assumption is appropriate. Since our design methods leverage sampling distribution segments, this better performance does not come with a substantial computational cost.

\bibliographystyle{chicago}